\documentclass[aps,prx,nolongbibliography,notitlepage,twocolumn,groupedaddress]{revtex4-2}
\usepackage{siunitx}
\usepackage{amsmath}
\usepackage{amssymb}
\usepackage{bbold}
\usepackage{color}
\usepackage{comment}
\usepackage{tikz}
\usepackage{pgfplots}
\pgfplotsset{compat=1.5}
\usepackage{graphicx}
\usepackage[colorlinks=true, linkcolor=black, citecolor=blue, pdfencoding=auto]{hyperref}
\usepackage{blindtext}
\usepackage{physics}

\newcommand{\intBZ}{ \int_{\rm BZ}  } 

\def\phdag{{\phantom \dagger}}

\newcommand{\rev}[1]{ {\color{black} #1}}

\begin{document}
\title{Instantaneous Response and Quantum Geometry of Insulators}

\author{Nishchhal \surname{Verma} }
\affiliation{Department of Physics, Columbia University, New York, NY 10027, USA}

\author{Raquel \surname{Queiroz} }
\email{raquel.queiroz@columbia.edu}
\affiliation{Department of Physics, Columbia University, New York, NY 10027, USA}
\affiliation{Center for Computational Quantum Physics, Flatiron Institute, New York, New York 10010, USA}

\begin{abstract}
We present the time-dependent Quantum Geometric Tensor (tQGT) as a comprehensive tool for capturing the geometric character of insulators observable within linear response. 
We show that tQGT describes the zero-point motion of bound electrons and acts as a generating function for generalized sum rules of electronic conductivity. It therefore enables a systematic framework for computing the instantaneous response of insulators, including optical mass, orbital angular momentum, and dielectric constant. This construction guarantees a consistent approximation across these quantities upon restricting the number of occupied and unoccupied states in a low-energy description of an infinite quantum system.
We outline how quantum geometry can be generated in periodic systems by lattice interference and examine spectral weight transfer from small frequencies to high frequencies by creating geometrically frustrated flat bands.
\end{abstract}

\maketitle

The groundbreaking result of Thouless–Kohmoto–Nightingale–den Nijs (TKNN) established that the quantization of Hall conductivity originates from a topological invariant associated with the phase winding of the ground-state wavefunction of an insulator \cite{TKNN1982}. This discovery made it clear that linear response cannot be fully characterized by energy dispersion or charge density alone. Instead, the phases of wavefunctions are essential ingredients in capturing transport phenomena of quantum materials. Over the past decade, the study of topological invariants encoded in wavefunction phases has become commonplace, and their significance for physical properties of materials is now widely recognized \cite{Vanderbilt2018Book, Hasan2010, Qi2011, Resta2011, Bradlyn2017, vergniory2019complete}.

The ubiquity of geometric contributions in linear \cite{Nagaosa2010, Xiao2010}, and nonlinear \cite{Lapa2019, Gao2019, Kozii2021}, response is rooted in the fact that electromagnetic fields couple to materials via the bulk dipole moment. Since the dipole moment itself is tied to the Berry phases of wavefunctions in insulators \cite{KingSmith1993}, the resulting response inevitably carries geometric information. This principle extends well beyond polarization and it is reflected in orbital magnetic moments \cite{Thonhauser2005, Xiao2005}, dielectric permittivity \cite{komissarov2023}, and superfluid stiffness and spectral weight in superconductors \cite{TormaNatPhy2015, Tovmasyan2016, Julku2016, Torma2021, Verma2021, Mao2023, Mao2024}. Nonlinear responses also exhibit similar structures, where higher powers of the dipole operator probe higher orders of geometry \cite{Ahn2022, Hetenyi2023}.  

A central mathematical object that unifies these observables is the quantum geometric tensor (QGT), $\mathcal{Q} = g + i \Omega / 2$, whose real and imaginary parts correspond to the quantum metric $g$ and the Berry curvature $\Omega$, respectively. The Berry curvature tracks the phase accumulated along closed paths in parameter space \cite{berry1984quantal}, while the quantum metric measures the loss of norm, or non-adiabaticity, in wavefunction evolution \cite{Provost1980, kolodrubetz2017geometry}. In crystalline systems, Berry curvature is often constrained to vanish by inversion and time-reversal symmetries, but the quantum metric is positive semi-definite and generally non-vanishing whenever more than one band is present. This makes the quantum metric especially relevant to both trivial and topological materials. A pressing challenge, however, is to identify physical observables that directly measure the quantum metric \cite{Torma2023commentary}. 
\rev{Although several proposals have been made \cite{Neupert2013, Miaomiao2023, rossi2021quantum, Verma2023, yu2023, Iskin2023, Chen2024, verma2024stepresponse, Mitscherling2022}, the metric usually appears entangled with energy prefactors, making direct isolation difficult, except in special cases such as Landau levels or flat bands \cite{Torma2021} or ultracold atoms \cite{asteria2019measuring, brown2022direct}.

An important path toward addressing this challenge lies in optical sum rules, which connect integrals of the AC conductivity to equilibrium geometric quantities. The Souza–Wilkens–Martin (SWM) sum rule \cite{Souza2000} is the only exact relation known to involve the metric directly and independently of band energies, stating that the first negative moment of the dynamical conductivity in insulators equals the ground-state quantum metric, or localization tensor \cite{Kohn1964,Resta2011}. Other optical sum rules highlight complementary aspects of geometry, the $f$-sum rule controls the plasma frequency and effective mass of the electronic degrees of freedom and in certain flat-band superconductors incorporates the quantum metric of the bands \cite{Verma2021, Mao2023}; the orbital magnetic moment is closely related to but distinct from the Berry curvature \cite{Thonhauser2005, Xiao2005, Xiao2010, Mahon2023}. These examples indicate that geometry pervades response functions, yet the precise role of geometry across different sum rules, and whether they can be organized within a single unifying principle, remains unclear. It has even become common to label any property influenced by wavefunctions as ``quantum geometric'', clearly hinting at the lack of a unifying principle.

To address this issue, we return to the fundamental observation that quantum geometry originates from projecting the position operator into the occupied subspace \cite{Marzari1997, Marzari2012}. From this perspective, it is natural to introduce the off-diagonal dipole operator $d = e\mathcal{D}$, with $\mathcal{D}(t) \equiv \hat{Q} \hat{r}_\mu(t)\hat{P}$, where $\hat{P}$ projects onto the occupied states and $\hat{Q} = 1-\hat{P}$ projects onto the unoccupied states. Its zero-temperature correlation function defines the time-dependent quantum geometric tensor (tQGT)
\begin{equation}
    \mathcal{Q}_{\mu\nu}(t-t') = \langle \mathcal{D}^\dagger_\mu(t)\mathcal{D}_\nu(t')\rangle = {\rm tr}\!\left[\hat{P}\,\hat{r}_\mu(t)\hat{Q}\,\hat{r}_\nu(t')\right]. \label{eq:mt:def:Qmunu}
\end{equation}

The tQGT captures zero-point motion in quantum systems: it vanishes for a classical particle, reduces to the localization tensor at $t=0$, and encodes both the metric and Chern number as its real and imaginary parts. 
Crucially, the inclusion of the projector $\hat{Q}$ eliminates gauge dependence, making tQGT manifestly gauge invariant and physically observable.
It circumvents the gauge-dependent singularities inherent to dipole–dipole correlation functions in the theoretical description of solids.

At finite times, the tQGT is not Hermitian but can be separated into a Hermitian $\mathcal{Q}_{\mu\nu}(t) + \mathcal{Q}_{\nu\mu}(-t)$ and an anti-Hermitian $(\mathcal{Q}_{\mu\nu}(t) - \mathcal{Q}_{\nu\mu}(-t))$ part, of which the former is tied to step response \cite{verma2024stepresponse} and the latter relates to the optical conductivity tensor.
We further note that notions of time-dependent quantum geometric tensor already exist for Hamiltonian dynamics along a path in parameter space \cite{rattacaso2020quantum, lu2010operator, Bleu2018, austrich2022quantum}. Our formalism is different as it introduces time dependence only in the operator $\mathcal{D}_\mu(t)$ while keeping the Hamiltonian time independent.

It is well known for isolated atoms interacting with a background gauge field that dipole transitions lead to vacuum fluctuations \cite{carmichael1999two}. As we highlight in Sec.~3, the tQGT provides a natural way to extend this concept to lattice systems and clarifies the origin of quantum geometry that was recently emphasized in flat-band superconductivity \cite{Torma2021}. By studying zero-point motion \cite{Resta2006}, one finds that quantities such as the orbital magnetic moment, the $f$-sum rule, and the dielectric permittivity all stem from the same projected dipole correlations. This observation motivates our central question: can all optical sum rules be unified as time derivatives of a single geometric object, and what new physical insights follow from such a unification? In this manuscript we show that the answer is yes. By deriving an explicit relation between the conductivity tensor and the time-dependent quantum geometric tensor, we establish a framework in which generalized sum rules arise systematically from its time derivatives. Within this framework, all generalized sum rules appear on equal footing, all uniquely fixed by the projection operator.
}

\section{Results}
Our main results can be summarized in a rewriting of the Kubo formula for conductivity in terms of tQGT in the time domain, and the consequent generalized form for the dissipative sum rules that tie various geometric properties of an insulating system.
To concisely isolate the dissipative response in conductivity, we define the axial vectors $\sigma^L_\mu$ and  $\sigma^H_\lambda$ to write the conductivity tensor as
\begin{equation}
    \sigma_{\mu\nu} = \delta_{\mu\nu} \sigma_{L,\mu} + \epsilon_{\mu\nu\lambda} \sigma_{H,\lambda}
\end{equation}
where $\mu,\nu,\lambda$ are spatial indices and the superscripts $L$ and $H$ refer to Longitudinal and Hall responses (see SI Appendix 5 for details). \rev{Here, no summation is implied over $\mu$ in the first term, while the index $\lambda$ in the second term is summed over according to the Einstein convention.
Our formalism is general and applies to arbitrary many-body quantum systems.}
In particular, for gapped quantum systems with charge conservation (see SI Appendix 1), we can write the conductivity in a spectral representation in terms of the matrix elements of the position operator
\begin{equation}
    \sigma_{\mu\nu}(\omega) = -i \dfrac{e^2}{\hbar}\sum\limits_{m\neq n} f_{nm} \omega_{nm} \hat{r}^{nm}_\mu \hat{r}^{mn}_\nu \dfrac{1}{\omega - \omega_{mn} }\label{eq:conductivityspectral}
\end{equation}
where $m,n$ are the energy states, $f_{nm} = f_n - f_m$ are the occupation factors with $\omega_{mn} = \omega_m - \omega_n$ being energy difference (see SI Appendix 1 for details). 
The formula points to generalized sum rules for the dissipative part of conductivity (see SI Appendix 1A for details)
\begin{equation}
     \mathcal{S}_{\mu\nu}^\eta = \int\limits_0^\infty d\omega \; \dfrac{ \sigma^{\rm abs}_{\mu\nu}(\omega) }{ \omega^{1-\eta} } = \delta_{\mu\nu} \mathcal{S}^\eta_{L,\mu} + i \epsilon_{\mu\nu\lambda} \mathcal{S}_{H,\lambda}^\eta.
\end{equation}
The absorptive (or Hermitian) part is defined as $\sigma^{\rm abs}_{\mu\nu} = (\sigma_{\mu\nu} + \sigma_{\nu\mu}^*)/2$ (see SI Appendix 5 for details).
The half limit is crucial for differentiating the sum rule from standard Kramers-Kronig relations (see SI Appendix 6).
Notice that $\sigma^{\rm abs}_{\mu\nu}(\omega)$ includes the real part of longitudinal conductivity ${\rm Re}[ \sigma_{L,\mu}(\omega) ]$ and the imaginary part of Hall conductivity ${\rm Im}[ \sigma_{H,\lambda}(\omega) ]$.
Eq.~\eqref{eq:conductivityspectral} bears some similarities with the tQGT, which in the same representation is given by 
\begin{equation}
    \mathcal{Q}_{\mu\nu}(t - t^\prime) = \sum\limits_{m, n} f_n(1-f_m) e^{i \omega_{mn} t} \; \hat{r}^\mu_{nm} \hat{r}^\nu_{mn} \label{eq:def:introduce-Q-munu-t:main}
\end{equation}
If we choose to write the conductivity in the time domain (see SI Appendix 2 for details), we indeed find that it assumes a remarkably simple form
\begin{equation}
    \sigma_{\mu\nu}(t) = \dfrac{\pi e^2}{\hbar} \;\Theta(t)\; \partial_t \mathcal{Q}^{\rm as}_{\mu\nu}(t).\label{eq:timedomaincond}
\end{equation}
where $\mathcal{Q}^{\rm as}_{\mu\nu}(t) \!=\! ( \mathcal{Q}_{\mu\nu}(t) - \mathcal{Q}_{\nu\mu}(-t))/i$ is the anti-symmetric and anti-Hermitian part of the tQGT (see SI Appendix 3 for details). Note that Eq.~\eqref{eq:timedomaincond} is valid for both metals and insulators, although in this work we focus on the latter, see a more detailed discussion in SI Appendix 4C. The relation between dynamics of electrons and conductivity is now evident, for insulators both $\mathcal{Q}_{\mu\nu}(t)$ and $\sigma_{\mu\nu}(t)$ have bounded oscillations that occur precisely because of quantum geometry.

\begin{figure}
    \centering
    \includegraphics[width=\columnwidth]{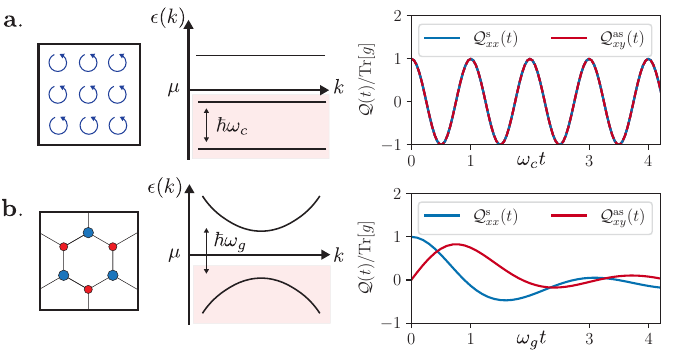}
    \caption{
    {\bf a}. Time-dependent quantum geometric tensor (tQGT) in Landau levels. The real longitudinal and imaginary Hall parts of the tQGT are identical and oscillate with the same frequency $\omega_c = \sqrt{e B/m}$. {\bf b}. tQGT in a honeycomb lattice model with nearest neighbor hopping and a $C_{2z}$ breaking mass term $m_z>0$ (details in Appendix I1). Different parts of tQGT now have different time profiles. As a result, geometric quantities arising from different derivatives are all distinct.
    }
    \label{fig1:mt}
\end{figure}

With this insight, it is straightforward to find an exact expression for all optical sum rules
\begin{equation}
   \mathcal{S}_{L/H,\mu}^\eta = \dfrac{\pi e^2}{\hbar} \left[ (-i \hat{\partial}_t)^\eta \mathcal{Q}_{L/H,\mu}(t) \right]_{t=0} \label{eq:main-result}
\end{equation}
where $\mathcal{Q}_{\mu\nu}(t)$ has been decomposed into longitudinal and Hall parts $\mathcal{Q}_{\mu\nu}(t) = \delta_{\mu\nu}\mathcal{Q}_{L,\mu}(t) + \epsilon_{\mu\nu\lambda}\mathcal{Q}_{H,\lambda}(t)$ for a succinct presentation.
This is the main result of our work. Each sum rule ($\mathcal{S}^\eta$) can be associated with an instantaneous property characterizing the bound electrons of the insulator. This instantaneous response, given by the various derivatives of $\mathcal{Q}(t)$ at $t=0$, defines various measurable properties associated with the zero-point motion of the bound charges. 
In particular, $\eta=0$ is the SWM sum rule for insulators \cite{Souza2000} and $\eta=1$ is the $f$-sum rule that defines the plasma frequency $\omega_p^2 = 4\pi n e^2/m$ \cite{PinesNozieres1989}.
\rev{Crucially, all of these sum rules follow from a single gauge-invariant quantity for Bloch electrons, providing a unified formulation that avoids the gauge-dependent artifacts of the length gauge and the diamagnetic correction required in the velocity gauge of light-matter coupling \cite{kobe1978gauge}.}

% SWM sum rule
We begin by unpacking Eq.~\eqref{eq:main-result} for various $\eta$.
The longitudinal and Hall sum rules take the explicit form in insulators
\begin{equation}
    \mathcal{S}_{L,\mu}^\eta = \dfrac{\pi e^2}{\hbar} \!\!\sum\limits_{ \substack{n\in {\rm filled}\\ m \in {\rm empty} } }\!\!\!\! \omega_{mn}^\eta g_\mu^{nm}, \!\quad \mathcal{S}_{H,\lambda}^\eta = \dfrac{\pi e^2}{\hbar} \!\!\sum\limits_{ \substack{n\in {\rm filled}\\ m \in {\rm empty} } }\!\!\!\! \omega_{mn}^\eta \Omega_\lambda^{nm} \label{eq:sumrule-S-eta}
\end{equation}
where $g_\mu^{nm} \equiv |\langle u_n | \hat{r}_\mu | u_m \rangle|^2$, $\Omega_\lambda^{nm} \equiv 2{\rm Im}[\hat{\lambda}\cdot\langle u_n | \hat{ {\bf r} } | u_m \rangle \times \langle u_m | \hat{{\bf r}} | u_n \rangle]$ are various matrix elements of the position operator and $\omega_{mn}$ is the energy difference between states $|u_m\rangle$ and $|u_n\rangle$.
These matrix quantities are all gauge invariant under a $U(1)$ transformation that gives arbitrary phases to a band wavefunction for $m\neq n$.
The case with $m=n$, which arises in metals, needs a more careful treatment that is outlined in SI Appendix 4C.

The $\eta=0$ sum rule has no energy pre-factor. Its longitudinal part is the SWM sum rule and defines the quantum metric $g$ and its Hall part can be obtained from a Kramers-Kronig relation of the TKNN conductivity. It is simply the total Chern number $\mathcal{C}$ of the system.
The $\eta=1$ sum rule defines the plasma frequency with $\omega_p^2 = 4\pi ne^2/m$ where $n$ is the electron total density (including core electrons) and $m$ is the bare mass of the electron, and from the Hall part one can read the orbital magnetic moment of insulators as defined in Refs.\cite{Souza2008, Resta2020}.
We should pause here to comment that in \emph{ab-initio} calculations, which start with free electrons subjected to a periodic potential, the number of bands is infinite. The plasma frequency is independent of the periodic potential and the exact form of the wavefunctions and it is guaranteed to be $\propto n/m$ by the effective band mass theorem \cite{ashcroft1977solid}. 
Deviations can occur for low-energy effective theories. For example, in a tight-binding model, especially those with multiple bands, the plasma frequency is renormalized \cite{Hazra2019, Bellomia2020} and becomes sensitive to the wavefunctions \cite{Iskin2019}. 
Finally, the $\eta=-1$ sum rule defines the dielectric permittivity of the material \cite{komissarov2023}.
Intuitively, it measures how much the electrons can polarize the medium in the presence of an electric field. 
The imaginary part of the same sum measures if electrons have a torsional twist in responding to an external electric field, reminiscent of the gyration vector \cite{Landau1984}.

The power of Eq.~\eqref{eq:sumrule-S-eta} lies in a consistent definition of the geometric properties of materials. Different sum rules are not independent, and in fact become fixed once the projectors $\hat{P}$ and $\hat{Q}$ are fixed, as well as the band energies. Therefore, we present a consistent definition of instantaneous properties for effective tight-binding models obtained by selecting a subset of bands obtained by first principles.
This subtlety has been especially relevant since the discovery of topological insulators. 
Topological indices act as an obstruction to finding a local basis that only has low-energy orbitals \cite{Monaco2018} while respecting all the spatial and internal symmetries \cite{SoluyanovPRB2011, SoluyanovPRB2012}.
To describe topological phases with local Hamiltonians it is necessary to include multiple orbitals which mandates bounded oscillations in the tQGT.

To gain some familiarity with the tQGT, it is instructive to consider Landau levels in the two-dimensional electron gas as the simplest example of a topological insulator. In this case it takes the simple form with $\mathcal{Q}(t) = \mathcal{C} e^{i \omega_c t} $ where $\omega_c = \sqrt{eB/m}$ is cyclotron frequency and $\mathcal{C}$ is the total Chern number.
The sum rules can be immediately deduced
\begin{equation}
\mathcal{S}^\eta_{\mu\nu} = \mathcal{C} \omega_c^\eta ( \delta_{\mu\nu} + i \epsilon_{\mu\nu z}) \label{eq:Sum-Rule-LL}
\end{equation}
where the area of the magnetic unit cell ($\ell_B^2 = eB/m$) cancels the volume normalization factor to give rise to the net Chern number $\mathcal{C}$.
The precise analytical form of the Landau level wavefunction enforces quantum metric, Berry curvature, orbital magnetic moment, effective optical mass, and dielectric permittivity -- all to be the same, up to factors of $\omega_c$.
This is consistent with the classical picture of electrons going around in cyclotron orbits in the plane perpendicular to the magnetic field.
The challenge lies in extrapolating these concepts over to real materials that have non-flat dispersions and therefore nontrivial dynamics in the tQGT.
Moreover, the wavefunctions themselves do not admit the so-called ideal band geometry \cite{Roy2014, Ledwith2023}.
As a result, all the geometric quantities are different, albeit related. 
\rev{Here, we find that sum rules allow us to straightforwardly compute these quantities, and that defining these moments provides a systematic framework for characterizing quantum materials.}

\section{Model and Methods}
Optical conductivity is the static $({\bf q}=0)$ response of a system to a dynamical $(\omega\neq 0)$ field. 
The Kubo formula for conductivity in the exact many-body basis can be written formally as
\begin{equation}
    \sigma_{\mu\nu}(\omega) = \dfrac{i}{\omega}( D_{\mu\nu} - \chi_{j_\mu,j_\nu}(\omega) )
\end{equation}
where $D_{\mu\nu} \equiv \langle \partial_{A_\mu A_\nu} \mathcal{H}(A) \rangle - \chi_{j_\mu,j_\nu}(0)$ 
is the charge stiffness defined in terms of diamagnetic current and paramagnetic current-current correlator with $j_\mu \equiv \partial_{A_\mu} \mathcal{H}(A)$ \cite{Allen2015}.
The charge stiffness can be equivalently defined as the change in the ground state energy of the system following a twist in the boundary conditions \cite{Kohn1964}. 
It vanishes identically for a gapped quantum system with number conservation \cite{scalapino1993}.
Considering such gapped quantum systems for now, we arrive at the Kubo formula for conductivity shown in Eq.~Eq.~\eqref{eq:conductivityspectral}. 
We focus on the dissipative component that arises from \rev{the imaginary part of the frequency denominator ${\rm Im}[1/(\omega - \omega_{mn})] = \pi\delta(\omega - \omega_{mn})$}. 
In two dimensions, it corresponds to ${\rm Re}[\sigma_{xx}]$ and ${\rm Im}[\sigma_{xy}]$. 
The three dimensional generalization is straightforward with the axial representation $\Omega_{\mu\nu}^{mn} \equiv \epsilon_{\mu\nu\lambda} \Omega_\lambda^{mn}$ that is outlined in Appendix F.
We rewrite absorptive part of the conductivity as 
\begin{equation}
\sigma^{\rm abs}_{\mu\nu}(\omega) = \delta_{\mu\nu} {\rm Re}[\sigma^L_\mu] + i\epsilon_{\mu\nu\lambda} {\rm Im}[\sigma^H_\lambda].
\end{equation}

The inclusion of the position operator in the Kubo formula is the key to extracting the quantum geometry.
Consider the longitudinal part for instance
\begin{equation}
    {\rm Re}[\sigma_{L,x}(\omega)] =  \dfrac{\pi e^2}{\hbar}\sum\limits_{m\neq n} f_{nm} \omega_{nm} g^{nm}_{L,\mu} \delta(\omega - \omega_{mn})
\end{equation}
where the \rev{quantum metric ($g_{L,\mu} \equiv \delta_{\mu\nu}g_{\mu\nu}$)} is nearly present but the pre-factor of $\omega_{mn}$ and the delta function prohibit $g_{xx}^{nm}$ from becoming the quantum metric.
One way to circumvent the issue is to consider sum rules that remove the delta function by
\begin{equation}
    \int\limits_0^\infty d\omega\; \delta(\omega - \omega_{mn} ) = \dfrac{\Theta(\omega_{mn})}{2}.
\end{equation}
As we show next, different moments with $\omega^{\eta-1}$ then correspond to different geometric properties of the insulator, arising from the time-dependence of the zero point motion.
\rev{Fig.~\ref{fig2} illustrates these sum rules for the Haldane model of a Chern insulator at half-filling across the topological phase transition. They track how bound electrons reorganize between trivial and topological insulating states, and notably the Hall sum rules for $\eta \leq -1$ exhibit an antisymmetric divergence at the transition.}

\subsection{SWM Sum Rule and Quantum Geometry}

The initial proposal behind the SWM sum rule \cite{Souza2000} was aimed at classifying metallic and insulating states by examining if the ground state quantum metric \rev{$g$} diverges in the thermodynamic limit \cite{Resta2011}, a line of reasoning that traces back to Kohn \cite{Kohn1964}.
The value of the metric itself was not considered important provided it remained finite. 
\rev{However, with the refined classification of insulators into atomic, obstructed, or topological \cite{Bradlyn2017}, revisiting the SWM sum rule \cite{Souza2000} has elaborated the connections between quantum geometry and sum rules \cite{onishi2023quantum, onishi2024quantum, onishi2024universal, ghosh2024probing, Kruchkov2023}.
} 

In our notation, the sum rule takes the expression
\begin{equation}
    \int\limits_0^\infty d\omega \; \dfrac{ \sigma^{\rm abs}_{\mu\nu}(\omega) }{ \omega } = \dfrac{\pi e^2}{\hbar} \big( \delta_{\mu\nu} g_{\mu\nu} + i\epsilon_{\mu\nu\lambda} \mathcal{C}_\lambda \big)
\end{equation}
whose real part is the SWM sum rule with ground state quantum metric $g_{\mu\nu}$ and the imaginary part is the Chern number $\mathcal{C}$.
We note that Chern number in ${\rm Im}[\sigma_H(\omega)]$ sum rule is related to ${\rm Re}[\sigma_{\lambda}^H(0)]$ as a consequence of Kramers–Kronig relations.

\rev{Let us briefly comment on the distinction between the quantum metric that appears in the SWM sum rule and the quantum metric that is defined for Bloch states of a single band.
It is apparent from the definition that tQGT depends sensitively on the choice of projector $P$. For non-interacting fermions in a periodic potential, this projector can be further resolved into bands \rev{$m$ and momentum $\bf k$}, $\hat{P}_m({\bf k}) = | \psi_{m, \bf{k}}\rangle \langle \psi_{m, \bf {k}} |$ where $| \psi_{m, \bf{k}}\rangle$ is  the Bloch wavefunction.
The projector into the ground state $\hat P$ is then the sum of $\hat{P}_m$ for all filled bands $m$ integrated over the Brillouin zone. It is often convenient to discuss the geometry of an individual band, with important consequences for superfluid stiffness in flat band superconductors \cite{Torma2021}. In this case, the formalism of the tQGT can be extended by substituting the ground state projector $\hat P$ to the band and momentum resolved one $\hat P_{m}(\bf k)$ in Eq.~\eqref{eq:mt:def:Qmunu}
Importantly, the QGT for the full projected space cannot be expressed as a sum of QGT of individual bands \cite{Mera2022}. The two quantities are thus different physical objects.}

\subsection{$f$-Sum Rule and Effective Mass}
The $f$-sum rule is given by the real part of the longitudinal conductivity
\begin{equation}
     \int\limits_0^\infty d\omega \; {\rm Re}[ \sigma_{L,\mu}(\omega) ]  = \dfrac{\pi e^2}{\hbar} \sum\limits_{ \substack{n\in {\rm filled}\\ m \in {\rm empty} } } \omega_{mn} g_{L,\mu}^{nm} \label{eq:mt:f-sum-rule}
\end{equation}
and is used to define the plasma frequency $\omega_p^2$ which involves the total electron density and bare electron mass. 

\begin{figure}
    \centering
    \includegraphics[width=\columnwidth]{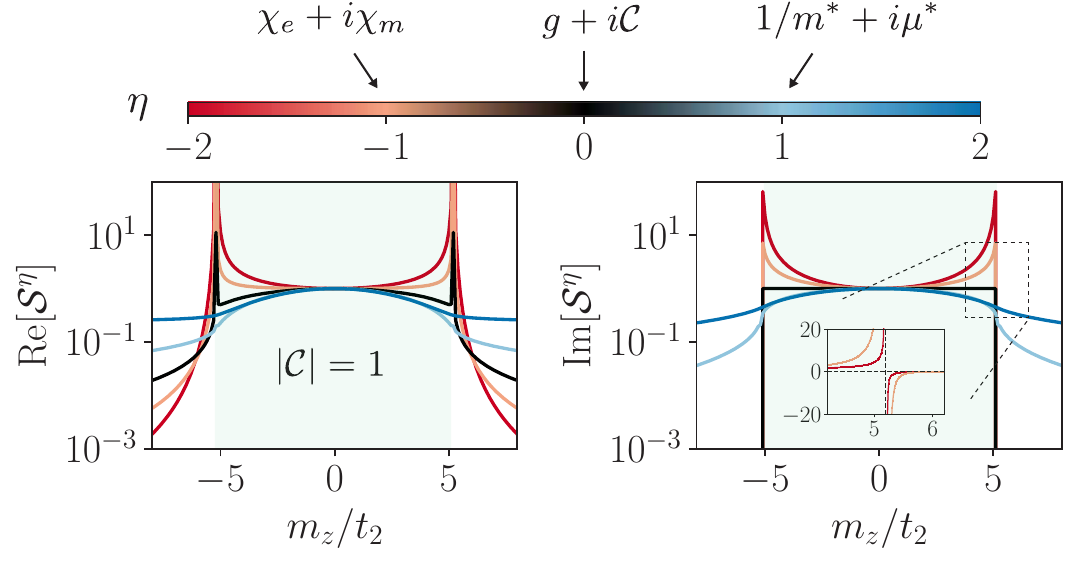}
    \caption{Longitudinal and Hall sum rules $(\mathcal{S}^\eta = \mathcal{S}_{L}^\eta+i \mathcal{S}_{H}^\eta)$ for the two-dimensional Haldane Chern Insulator at half-filling across the topological phase transition that occurs at $|m_z/t_2| = 3\sqrt{3}$.
    The model includes a nearest neighbor hopping $t$, a next nearest neighbor hopping $t_2$ with flux $\phi = \pi/2$, and an inversion breaking mass $m_z$ (see Appendix I2 for details).
    The sum rules are normalized by their value at $m_z=0$. The SWM sum rule (black) in the left panel corresponds to ground state quantum metric $g$ and the Chern number $\mathcal{C}$. The metric diverges across the topological phase transition. The discontinuities become more smooth or divergent depending on the energy power $\eta$. Note that the Hall sum rules with $\eta \leq -1$ show an anti-symmetric divergence at the phase transition.
    }
    \label{fig2}
\end{figure}

In crystalline solids, the label for each quantum states breaks into band label $m$ and crystal momentum label ${\bf k}$. As we outline in Appendix E1, the algebra is identical except for the additional label and integral over ${\bf k}$.
The plasma frequency in such systems is typically a large energy scale that includes all possible ${\bf q}=0$ direct inter-band transitions that are allowed in ${\rm Re}[\sigma_{L,\mu}]$.

The mass obtained from the sum rule deviates from the bare electron mass when we use low-energy sum rules that include transitions only between the low-energy orbitals.
As the Hilbert space is truncated, for instance in tight-binding models, the band mass theorem no longer applies and the sum rule depends sensitively on wavefunctions.
By requiring the low-energy plasma frequency to be proportional to $n/m$, we can define an effective mass for each filled band:
\begin{equation}
    \mathcal{M}_{n,\mu}^{-1} \equiv \sum_{m\neq n}\omega_{mn} g_\mu^{mn},
\end{equation}
such that the sum rule is $\mathcal{S}_{L,\mu}^1 = \sum_{n \in {\rm filled}} \mathcal{M}_{n,\mu}^{-1} \equiv n/m_g^*$ with $m_g^*$ being the total geometric contribution to the effective mass.
We emphasize that this is the only consistent way of defining optical mass in topological insulators.

The theory extends easily to metals that are gapless and have a Fermi surface Drude contribution. 
As we show in Appendix E, the existence of Fermi-surface adds an unbounded linear-in-time contribution to the tQGT
\begin{equation}
    \mathcal{Q}_{\mu\nu}(t) = D_{\mu\nu} t + \mathcal{Q}^{\rm inter}_{\mu\nu}(t) + \cdots
\end{equation}
where $(\cdots)$ includes a time-independent piece that does not contribute to the $f$-sum rule (see Appendix 4B for details).
The Drude weight $D_{\mu\nu} = \sum_{n,{\bf k}} f_{n, {\bf k} } \partial_k^2 \epsilon_{n,{\bf k}}$ contributes only to the longitudinal conductivity.
It adds a mass to the $f$-sum rule that is given by the band curvature, $m^*_{\epsilon}$. In total, we get
\begin{equation}
    {1\over m^*}={1\over m_\epsilon^*}+{1\over m_g^*}
\end{equation}
such that the optical mass includes the geometric contribution and Fermi surface contributions in equal footing.
It can be regarded as an effective band mass theorem for the low-energy states.
Interactions modify the sum rule, however, within Fermi-liquid theory, the low-energy sum rule is tied to the Luttinger invariant \cite{Seki2017,Kruchkov2023}.

The geometric mass plays a crucial role in flat band superconductivity \cite{Torma2021}.
Without interactions, a flat band has an infinite density of states and a vanishing Fermi velocity.
The band curvature is infinite, $1/m^*_\epsilon=0$, but the geometric mass $m_g^*$ is finite.
If the band is further isolated from the rest of the spectrum, the system is insulating and the associated tQGT has bounded oscillations.
As the system becomes superconducting with an attractive Hubbard interaction $U$, the massively degenerate flat band manifold of states reorganizes into a lower Hubbard band with doubly occupied sites and an upper Hubbard band with single occupancy \cite{Tovmasyan2016,Verma2021}. With no single-particle bandwidth, the attractive interaction is the only energy scale in the problem.
The low-energy effective theory then corresponds to bosons with hopping amplitude set by the interaction times the quantum metric \cite{Tovmasyan2016, Herzog-Arbeitman2022}.

\rev{
At this point, it is important to highlight a subtlety with orbital embedding that arises in geometric mass.
tQGT, as defined in Eq.~Eq.~\eqref{eq:def:introduce-Q-munu-t:main}, depends sensitively on orbital embedding within the unit cell.
Of late, there has been growing interest around differentiating quantities that are geometry independent, like Chern number \cite{Simon2020}, and geometry dependent, like quantum metric. It is now understood that mean-field superfluid stiffness is in-fact geometry independent when one considers the generalized random phase approximation \cite{Huhtinen2022, Tam2024}. The sum rule, on the other hand, is geometry dependent. This does not present a contradiction, as geometric mass $m_g^*$ and mass of Cooper pair $m_b$ are two unrelated quantities. The only relation between the two comes from the $f$-sum rule being an upper bound on the stiffness. It is however interesting to note that geometric contributions make the $f$-sum rule finite and thus the stiffness is allowed to be finite.
}

The imaginary part of $\eta=1$ sum rule is also interesting. It corresponds to the dichroic sum rule \cite{Oppeneer1998} 
\begin{equation}
     \int\limits_0^\infty d\omega \; {\rm Im}[ \sigma_{H,\lambda}(\omega) ]  = \dfrac{\pi e^2}{\hbar} \sum\limits_{ \substack{n\in {\rm filled}\\ m \in {\rm empty} } } \omega_{mn} \Omega_\lambda^{nm}  \label{eq:mt:dichroic-rule}
\end{equation}
and defines the bulk orbital magnetic moment of the system.
This formula can be intuitively understood from a short time expansion of the tQGT in Eq.~Eq.~\eqref{eq:mt:def:Qmunu}
\begin{equation}
    \mathcal{Q}_{\mu\nu}(t) \approx  \mathcal{Q}_{\mu\nu}(0) + i t \tr [ \hat{P} [\hat{H},\hat{r}_\mu] \hat{Q} \hat{r}_\nu]
\end{equation}
where $[\hat{H}, \hat{r}_\mu] \equiv i \hbar \hat{v}_\mu$ is the velocity operator. The imaginary part of the first derivative then directly translates to an effective angular momentum $\approx \hat{z}\cdot {\rm Tr}[ \hat{P} {\bf v} \times \hat{Q} {\bf r}]$ from which the orbital magnetic moment can be derived \cite{Souza2008}.

\subsection{Permittivity tensor}
The $\eta=-1$ sum rule is defined only for insulators as
\begin{equation}
    \int\limits_0^\infty d\omega \; \dfrac{ \sigma^{\rm abs}_{\mu\nu}(\omega)  }{ \omega^2 } = \dfrac{\pi e^2}{\hbar} \big( \chi_{\mu}^e \delta_{\mu\nu} + i\epsilon_{\mu\nu\lambda} \chi_{\lambda}^m \big)
\end{equation}
where $\chi_{\mu}^e$ defines the longitudinal polarizability of the material in response to an electric field. It relates to the longitudinal dielectric constant $\epsilon=1+\chi^e$, and the geometric capacitance $c=\epsilon_0\chi^e$ of an insulator \cite{komissarov2023}. 
It quantifies the inverse spring constant of the electron that is tied to the atomic site. Here, $\epsilon_0$ is the vacuum permittivity.

The Hall part of the sum rule, $\chi_{\lambda}^m$, concerns imaginary part of the Hall conductivity ${\rm Im}[\sigma_H(\omega)]$ which is itself related to the real part of the dielectric function ${\rm Re}[\epsilon_H(\omega)]$.
If the integral limits were taken from $-\infty$ to $\infty$, the sum rule would evaluate to ${\rm Re}[\epsilon_H(0)]$ (by Kramers-Kronig relation) and vanish identically.
The sum rule survives because the limits are from $0$ to $\infty$.
Now since it is no longer a Kramers-Kronig relation, the sum rule does not correspond to an obvious zero-frequency property, similar to quantum metric ${\rm Tr}[g]$ and magnetic moment $\mu_\lambda$.
% need to plan carefully what to say next
Intuitively, ${\rm Im}[\epsilon_H(\omega)]$ characterises absorption of circularly polarized light and appears in dichroic responses.
Therefore, $\chi_{\lambda}^m$ is expected to captures the response of the electronic state to external fields of different polarization. 
Presumably it acts as a torsional constant, however, a precise definition is lacking.

\subsection{Exact Bounds on Sum Rules}
As we have emphasized in the previous section, different sum rules probe different instantaneous properties of the material.
These properties are independent but have certain constraints which we now outline.
Focusing on the longitudinal sum rule
\begin{equation}
    \mathcal{S}^\eta_{L,\mu} = \int\limits_{0}^\infty d\omega\; \dfrac{ {\rm Re}[\sigma_{L,\mu}(\omega)] }{\omega^{1-\eta}}
\end{equation}
we find a series of exact bounds
\begin{equation}
    \mathcal{S}^{m+n}_{L,\mu} \leq \sqrt{ \mathcal{S}^{2m}_{L,\mu} \;\mathcal{S}^{2n}_{L,\mu} }
\end{equation}
where $m,n$ are arbitrary real numbers (see Appendix H for details).
These bounds rely on the positive semi-definite property of ${\rm Re}[\sigma_{L,\mu}]$ \cite{traini1996electric} and hence do not work for the imaginary parts.
We also note that Landau levels trivially saturate the inequality.
As seen in Eq.~Eq.~\eqref{eq:Sum-Rule-LL}, all properties of Landau levels are one and the same upto factors of $\omega_c$.

\begin{figure}
    \centering
\includegraphics[width=0.95\columnwidth]{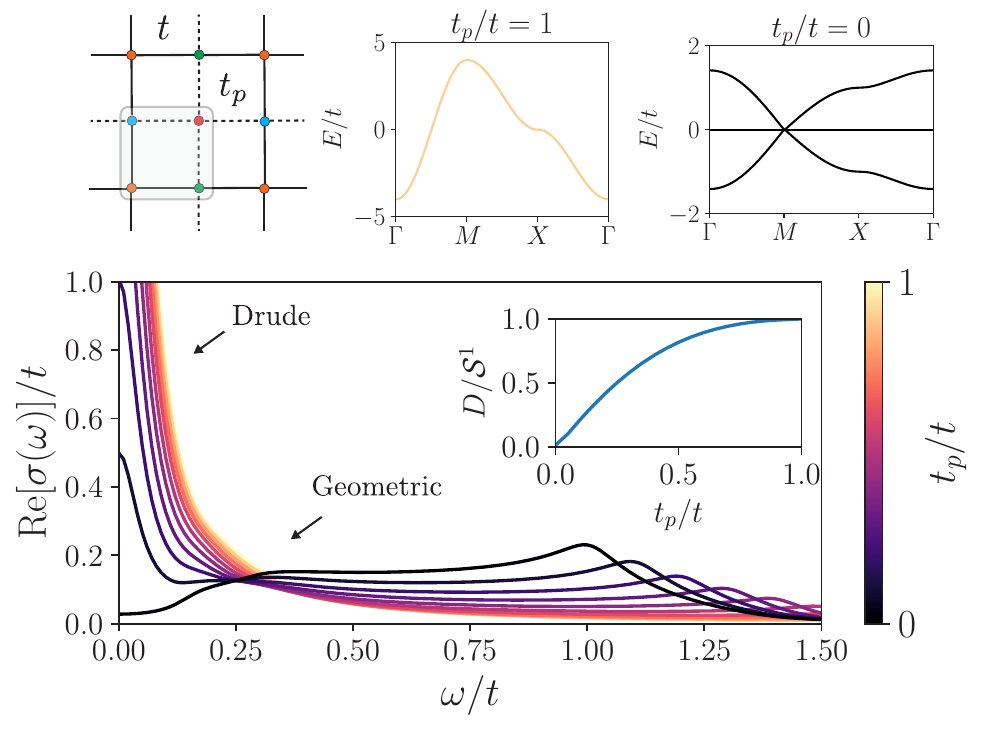}
    \caption{Transfer of Spectral Weight to higher frequencies in the Lieb lattice model that interpolates to the square lattice with parameter $t_p/t$. 
    At $t_p=t$, the square lattice has no geometric contribution and follows a Drude-like behavior. The inter-band transitions increase ss $t_p$ is reduced. Finally, at $t_p=0$, one of the orbital factors out and the resulting Lieb lattice has a vanishing Drude weight. The optical conductivity, on the other hand, has a non-zero interband contribution coming from the minimal inter-band conductivity of a Dirac cone \cite{Sun2018}.}
    \label{fig:fig3}
\end{figure}

The bound is most insightful for $m=-n= -1/2$ where it yields $\mathcal{S}^{0}_{L,\mu} \leq \sqrt{ \mathcal{S}^{1}_{L,\mu}\;\mathcal{S}^{-1}_{L,\mu} }$. In terms of material properties, it implies a universal inequality between quantum geometry, electric susceptibility and plasma frequency
\begin{equation}
    g \leq \omega_p \sqrt{ \chi^e }.
\end{equation}
The plasma frequency as written includes the bare mass of the electron and total density. 
It is a large number which can be modified, as outlined in the previous section, to get a tighter bound using a low-energy tight-binding description.
However, we emphasize that all sum rules should be defined consistently following the same tQGT. 
In other words, they should all agree on the  definition of the projector $\hat{P}$ and $\hat{Q}$.

Various sum rules are shown in Fig.~\ref{fig2} for the Haldane model of Chern insulator at half-filling across the topological phase transition. The sum rules track the reorganization of bound electrons in trivial and topological insulating states.
Interestingly, the Hall sum rules for $\eta \leq -1$ exhibit an anti-symmetric divergence at the transition.

\section{Inducing Dynamics in tQGT}

Two systems with identical spectrum can react differently to interactions due to difference in the wavefunctions. 
This variation is rooted in form factors that appear in projected interactions and can significantly influence the ground state of the system.
In our language, this fact is imprinted in the bounded oscillations of $\mathcal{Q}(t)$.
Understanding the origins of these oscillations, \rev{assessing their relevance in modeling quantum materials}, and determining when they might be negligible are some of the essential questions, particularly important in the context of moiré materials where flat bands emerge from an assembly of thousands of atoms \cite{Cao2018}.
In such materials, the nuanced effects of wavefunctions are expected to critically influence the correlated phenomena \cite{Classen2020}.

Let us investigate this effect in a simple toy model, where we illustrate the appearence of quantum geometry and oscillations in $\mathcal{Q}(t)$ with a single tuning parameter.  
We use a tight-binding model that interpolates between the Lieb lattice, which has a flat band, to the single band square lattice, which has no band geometry, by tuning $t_p/t$ (see Appendix I3 for details). Here $t$ is the hopping amplitude between nearest neighbors and $t_p$ connects the plaquette orbital to the rest of the lattice, see Fig.\ref{fig:fig3}.

Initially, at $t_p=t$, the Hamiltonian is simply an electron in a square lattice at half filling and therefore exhibits metallic behavior. It is characterized by a finite Drude peak $(\sigma(\omega) \propto D\delta(\omega))$ and a corresponding $\mathcal{Q}(t) \propto i D t$ leading to a pronounced peak at $\omega =0$ in dynamical conductivity.
As $t_p/t$ decreases, the plaquette orbital becomes more isolated, causing a reduction in the Drude weight.
This transition culminates in the formation of a Lieb lattice with a completely flat band at $t_p=0$ as observed in the dynamical conductivity ${\rm Re}[\sigma(\omega)]$
(Fig.\ref{fig:fig3}), where a shift of spectral weight towards higher frequencies accompanies the diminishing Drude weight. 
However, this redistribution of spectral weight only marginally affects the 
$f$-sum rule. 
In essence, the geometric frustration decreases the band curvature $1/m_\epsilon$ and at the same time increases the geometric contribution $1/m_g$.
The geometric contribution to the mass implies that there is an additional spectral weight that can flow back to zero frequency upon the inclusion of a local perturbation, for example, a local interaction, $U$.
In sum, the flat band in the Lieb lattice is more susceptible to spectral weight transfer than a conventional flat band.
A trivial atomic insulator will not feature this spectral weight transfer with interactions.
The enhanced susceptibility in principle can extend to the flat band in moire materials which are also born out of geometric frustration with thousands of atoms.

\section{Discussion}
\rev{The time-dependent Quantum Geometric Tensor offers a consistent, gauge-independent framework for defining geometric properties that arise within the linear response of gapped many-body quantum systems.}
The real and imaginary parts of its instantaneous ($t=0$) value determine the quantum metric and the Chern number, respectively. 
Through various time derivatives, the tQGT enables a consistent definition of optical mass, orbital moment, and dielectric constants of insulators, among other quantities left to explore.

In the time domain, the tQGT exhibits bounded oscillations that correspond to zero point motion of bound electrons, which is important for properly characterizing insulators.
Such oscillations are often overlooked in low-energy tight-binding models that are formulated based on \emph{ab-initio} band structures. 
Nonetheless, these oscillations may become significant whenever there is a discrepancy between low-energy and local basis. 
This scenario is common not only in topological insulators but also in geometrically frustrated lattices and possibly moire superlattices \cite{Julku2020, Hu2019, mendez2023theory}, where the spectral weight flows to higher frequencies and shows a tendency to revert upon incorporating interactions \cite{Xie2019, Herzog-Arbeitman2022, Torma2021, Xu2022}.

%============================================================

\section{Acknowledgements}
Work on quantum geometric properties of quantum materials is supported as part of Programmable Quantum Materials, an Energy Frontier Research Center funded by the U.S. Department of Energy (DOE), Office of Science, Basic Energy Sciences (BES), under award DE-SC0019443. 
We acknowledge discussions with Liang Fu and Yugo Onishi and thank them for discussing several aspects of their work; as well as Tobias Holder and Ilya Komissarov. \rev{We thank Jie Wang for pointing out the finite temperature generalization.}
The Flatiron Institute is a division of the Simons Foundation. 

\bibliography{tQGT}

\clearpage

\onecolumngrid
\newpage
\makeatletter 

\begin{center}
\textbf{\large Supplementary material for ``\@title ''}

Nishchhal Verma and Raquel Queiroz

\textit{Department of Physics, Columbia University, New York, NY 10027, USA}

\textit{Center for Computational Quantum Physics, Flatiron Institute, New York, New York 10010, USA} 

\end{center}
\vspace{20pt}

\setcounter{figure}{0}
\setcounter{section}{0}
\setcounter{equation}{0}

\renewcommand{\thefigure}{S\@arabic\c@figure}
\renewcommand{\thesection}{S-\@Roman\c@section}
\makeatother

\appendix

\rev{
\section*{Notation}
To ensure clarity in the supplementary material, we will use the following notation:
\begin{itemize}
    \item $m,n$ are labels for states with energies $E_m, E_n$ and states $|u_m\rangle, |u_n\rangle$
    \item $\mu,\nu$ are spatial indices
    \item $f_m$ are many-body occupation factors for many-body energy level $E_m$
    \begin{equation}
        f_m = \dfrac{1}{Z}e^{-\beta E_m}, \quad \text{where} \quad Z = \sum\limits_n e^{-\beta E_n}
    \end{equation}
    \item $\hat{\rho}$ is the thermal density matrix
    \begin{equation}
        \hat\rho = \exp(-\beta \mathcal{H}) = \sum\limits_m f_m |u_m \rangle \langle u_m |
    \end{equation}
    \item $\tr$ refers to trace over states
    \begin{equation}
        \left\langle \hat{O} \right\rangle = \tr[\hat{\rho} \hat{O} ] \equiv  \sum\limits_{m} f_m \langle u_m | \hat{O} | u_m \rangle = \sum\limits_{m} f_m O_{mm}
    \end{equation}
    for any operator $\hat{O}$
    \item $\hbar\omega_{mn}=E_m-E_n$ is the difference between energy levels $E_m$ and $E_n$
    \item $|u_{m, {\bf k} }\rangle$ are the cell-periodic parts of the Bloch wavefunction
    \item $f_m^0$ are the Fermi occupation factors 
    \begin{equation}
        f_{m, {\bf k}}^0 = \dfrac{1}{1 +  e^{ \beta(\epsilon_{m,{\bf k}} - \mu) }}
    \end{equation}
    \item $L/H$ refer to Longitudinal and Hall components
\end{itemize}
}
\section{Kubo Formula for Conductivity with Position Operators}\label{app:sec:Kubo-formula-cond}

We consider a general many-body Hamiltonian $\mathcal{H}$ in the presence of a uniform vector potential ${\bf A}$ with ${\bf q} =0$ and $\omega\neq 0$. Upon expanding the Hamiltonian around ${\bf A}=0$, we find the paramagnetic and diamagnetic current operators: 
\begin{equation}
    j_\mu = - \dfrac{\delta \mathcal{H}}{ \delta A_\mu} = j_{p,\mu} + A_\nu j_{d,\mu\nu}
\end{equation}
and use linear response theory to arrive at the Kubo formula for conductivity in the exact many-body basis
\begin{equation}
    \sigma_{\mu\nu}(\omega) = \dfrac{1}{i \omega} \left( \langle j_{d,\mu\nu} \rangle - \chi_{j_{p,\mu}, j_{p,\nu}  }(\omega) \right) 
\end{equation}
where $\langle j_{d,\mu\nu} \rangle$ is the diamagnetic response and $\chi_{j_{p,\mu}, j_{p,\nu}}(\omega)$ is the current-current correlator
\rev{
\begin{equation}
    \langle j_{d,\mu\nu} \rangle = \sum\limits_{m} f_m [ j_{d,\mu\nu} ]_{mm}, \quad \chi_{j_{p,\mu}, j_{p,\nu}}(\omega) = \sum\limits_{m,n} \dfrac{f_m - f_n}{ \omega - \omega_{mn}} [j_{p,\mu}]_{mn} [j_{p,\nu}]_{nm}\label{eq:def:chi-jj}
\end{equation}
where $f_m$ are the occupation factors, $\omega_{mn}$ are energy differences and $[j_{p,\mu}]_{mn} = \langle u_m | j_{p,\mu} | u_n \rangle$ are the matrix elements of the current operator. The zero-frequency limit of this correlator is subtle for gapless systems. To circumvent that, we combine the diamagnetic current and zero frequency part into the charge stiffness \cite{scalapino1993}
\begin{equation}
    D_{\mu\nu} = \langle j_{d,\mu\nu} \rangle - \lim\limits_{\omega\rightarrow 0} \chi_{j_{p,\mu}, j_{p,\nu}  }(\omega)
\end{equation}
which is finite only for systems with gapless electronic excitations.
The $\omega\rightarrow 0$ limit ensures that $m= n$ terms are excluded from the summation in zero frequency susceptibility.
Inserting this into the expression for conductivity, we find
\begin{equation}
    \sigma_{\mu\nu}(\omega) = \dfrac{1}{i \omega} \left( D_{\mu\nu} + \chi_{j_{p,\mu}, j_{p,\nu}  }(0) - \chi_{j_{p,\mu}, j_{p,\nu}  }(\omega) \right) = \dfrac{D_{\mu\nu}}{i \omega} + \dfrac{i}{\omega}( \chi_{j_{p,\mu}, j_{p,\nu}  }(\omega) - \chi_{j_{p,\mu}, j_{p,\nu}  }(0) )
\end{equation}
where the quantity in the second bracket can be manipulated as
\begin{equation}
    \dfrac{i}{\omega}( \chi_{j_{p,\mu}, j_{p,\nu}  }(\omega) - \chi_{j_{p,\mu}, j_{p,\nu}  }(0) ) = \dfrac{i}{\omega} \sum\limits_{m,n} \left( \dfrac{1}{\omega - \omega_{mn}} + \dfrac{1}{\omega_{mn}} \right) f_{mn} [j_{p,\mu}]_{mn} [j_{p,\nu}]_{nm} = i \sum\limits_{m,n} \dfrac{ f_{mn} }{ \omega_{mn} } [j_{p,\mu}]_{mn} [j_{p,\nu}]_{nm} \dfrac{1}{\omega - \omega_{mn}}
\end{equation}
which gives
\begin{equation}
    \sigma_{\mu\nu}(\omega) = i\dfrac{D_{\mu\nu}}{\omega} - i \dfrac{e^2}{\hbar}\sum\limits_{m\neq n} f_{nm} \omega_{nm} \hat{r}^{nm}_\mu \hat{r}^{mn}_\nu \dfrac{1}{\omega - \omega_{mn} } \label{app:eq:def-sigma}
\end{equation}
that is the same as before, except for the current operator swapped with position operator. The relation between matrix elements of the two operators comes from the definition
\begin{equation}
    j_\mu = i\dfrac{e}{\hbar} [ \mathcal{H}, \hat{r}_\mu ], \quad j_\mu^{mn} = i\dfrac{e}{\hbar} \omega_{mn} \hat{r}^{nm}_\mu , \quad m \neq n
\end{equation}
which holds as long as the eigenstates are different, $m\neq n$. 
Since we have organized the diagonal components into $D_{\mu\nu}$, this substitution works perfectly fine for the second term of Eq.~\eqref{app:eq:def-sigma}.}

We emphasize that the Eq.~\eqref{app:eq:def-sigma} is formally true even with interactions and disorder.
The main advantage of using position matrix elements instead of current becomes clear when we split the position vectors into symmetric and anti-symmetric parts
\begin{equation}
    \hat{r}^{nm}_\mu \hat{r}^{mn}_\nu = g_{\mu\nu}^{nm} + \dfrac{i}{2} \Omega^{nm}_{\mu\nu} \label{eq:g-omega-decomposition-of-xmu-xnu}
\end{equation}
with
\begin{equation}
    g_{\mu\nu}^{nm} = \dfrac{1}{2}( \hat{r}^{nm}_\mu \hat{r}^{mn}_\nu + \hat{r}^{nm}_\nu \hat{r}^{mn}_\mu  ), \quad i\Omega_{\mu\nu}^{nm} = \hat{r}^{nm}_\mu \hat{r}^{mn}_\nu - \hat{r}^{nm}_\nu \hat{r}^{mn}_\mu .
\end{equation}
Note that $g$ is symmetric under spatial indices $\mu\leftrightarrow \nu$ as well as eigenstates $n\leftrightarrow m$ while $\Omega$ is anti-symmetric in both.
The notation is chosen this way since the matrix elements are the precursors to the quantum metric and Berry curvature. We will show this explicitly for non-interacting electrons in Sec.~\ref{app:sec:tQGT-in-solids}.
The decomposition allows a succinct representation of conductivity that explicitly shows the quantum geometric content that is usually hidden in the position operators \rev{
\begin{equation}
    \sigma_{\mu\nu}(\omega) = i\dfrac{D_{\mu\nu}}{\omega} -i \dfrac{e^2}{\hbar}  \sum\limits_{m\neq n} f_{nm}  \omega_{nm}(g_{\mu\nu}^{nm} + i \Omega^{nm}_{\mu\nu}/2) \dfrac{1}{\omega - \omega_{mn} } \label{eq:sigma-munu-def}.
\end{equation}}
We note that although quantum geometry makes an appearance in the Kubo formula, extracting a precise observable that tracks it is not straightforward \cite{verma2024stepresponse}.

\subsection{Dissipative and Reactive Components in Insulators}\label{app:subsec:Dissipative}
\rev{Insulators have gapped electronic excitations and hence $D_{\mu\nu} = 0$. The conductivity is entirely inter-band. We next use the Sokhotski–Plemelj theorem}
\begin{equation}
    \dfrac{1}{\omega - \omega_{mn} } = \mathcal{P} \left[ \dfrac{1}{\omega - \omega_{mn}} \right] - i \pi \delta( \omega - \omega_{mn} ),
\end{equation}
where $\mathcal{P}$ denotes the Cauchy principle value, to rewrite conductivity in Eq.~\eqref{eq:sigma-munu-def}, as
\begin{equation}
    \sigma_{\mu\nu}(\omega) = -i \dfrac{e^2}{\hbar}  \sum\limits_{m\neq n} f_{nm} \omega_{nm} \; (g_{\mu\nu}^{nm} + i \Omega^{nm}_{\mu\nu}/2)  \; \left( \mathcal{P} \left[ \dfrac{1}{\omega - \omega_{mn}} \right] - i \pi \delta( \omega - \omega_{mn} )  \right)
\end{equation}
whose real and imaginary parts are
\begin{subequations}
\label{eq:sigma-dsp}
\begin{align}
    {\rm Re}[ \sigma_{\mu\nu}(\omega)] &= \dfrac{e^2}{2\hbar}  \sum\limits_{m\neq n} f_{nm} \omega_{mn} \left( 2\pi g_{\mu\nu}^{nm} \delta( \omega - \omega_{mn}) - \Omega^{nm}_{\mu\nu} \mathcal{P} \left[ \dfrac{1}{\omega - \omega_{mn}} \right]     \right) \label{eq:Re-sigma-dsp} \\
    {\rm Im}[\sigma_{\mu\nu}(\omega)] &= \dfrac{e^2}{2\hbar}  \sum\limits_{m\neq n} f_{nm} \omega_{mn} \left( \pi \Omega_{\mu\nu}^{nm} \delta( \omega - \omega_{mn}) - 2g^{nm}_{\mu\nu} \mathcal{P} \left[ \dfrac{1}{\omega - \omega_{mn}} \right]     \right). \label{eq:Im-sigma-dsp}
    \end{align}
\end{subequations}
The real and imaginary parts include both dissipative (with $\delta(x)$ functions) and reactive (with principle value) components.
\rev{Notice that the factor $f_{nm} \omega_{mn}$ is positive because $f$ is a decreasing function of energy, ensuring that the ${\rm Re}[\sigma_{xx}]$ is positive definite, which is necessary for the thermodynamic stability of the system.} Although our primary focus is on the dissipative parts and their corresponding sum rule, it is important to note that the reactive part leads to the well-known TKNN formula. In the dc limit, where $\omega =0$, we get
\begin{equation}
    {\rm Re}[\sigma_{\mu\nu}(0)] = \dfrac{e^2}{\hbar}  \sum\limits_{m\neq n} f_{nm} \Omega_{\mu\nu}^{mn}/2 \neq 0, \quad {\rm Im}[\sigma_{\mu\nu}(0)] = \dfrac{e^2}{2\hbar}  \sum\limits_{m\neq n} f_{nm} g_{\mu\nu}^{mn} = 0
\end{equation}
\rev{where the second part vanishes because $g_{\mu\nu}$ is symmetric in $m\leftrightarrow n$ while $f_{nm}$ is anti-symmetric.
Regarding the first term, we substitute the definition of $\Omega_{\mu\nu}$ to get
\begin{eqnarray}
  {\rm Re}[\sigma_{\mu\nu}(0)] = \dfrac{e^2}{\hbar}  \sum\limits_{m\neq n} f_{nm} (\hat{r}^{nm}_\mu \hat{r}^{mn}_\nu - \hat{r}^{nm}_\nu \hat{r}^{mn}_\mu) = \dfrac{e^2}{\hbar} \sum\limits_{m,n} f_n (\hat{r}^{nm}_\mu \hat{r}^{mn}_\nu - \hat{r}^{nm}_\nu \hat{r}^{mn}_\mu)
  = \dfrac{e^2}{\hbar} {\rm tr}(\rho [\hat{r}_\mu, \hat{r}_\nu]  ) = \dfrac{e^2}{h} \epsilon_{\mu\nu} \mathcal{C}
\end{eqnarray}
where the commutator is given by the Chern number $\mathcal{C} = {\rm tr}(\rho [\hat{r}_\mu, \hat{r}_\nu]  )/2\pi$ and $\epsilon_{\mu\nu}$ is the Levi-Civita tensor}.
This is the TKNN formula for Hall conductivity $\sigma_H = \sigma_{xy} - \sigma_{yx} = (e^2/h) \mathcal{C}$ \cite{TKNN1982}. It is the first instance where quantum geometry leads to an observable in conductivity.

\section{Conductivity of Insulators in time domain}\label{app:sec:sigma-time}
Intuitively, conductivity arises from the motion of charges. 
Since charges are frozen in insulators, the naive expectation is that insulators will have trivial dynamics.
It is however interesting to analyze the motion from a perspective of quantum geometry.
The conductivity in the time domain is the Fourier transform of the expression in frequency domain
\begin{eqnarray}
    \sigma_{\mu\nu}(t) = \int\limits_{-\infty}^\infty d\omega \; e^{i \omega t} \; \sigma_{\mu\nu}(\omega) &=&  \dfrac{2\pi e^2}{\hbar}  \sum\limits_{m\neq n} f_{nm} \omega_{nm} \; (g_{\mu\nu}^{nm} + i \Omega^{nm}_{\mu\nu}/2) \left[ \int\limits_{-\infty}^\infty \dfrac{d\omega}{2\pi i} \; \dfrac{e^{i \omega t}}{\omega - \omega_{mn}} \right]\\
    &=& \dfrac{2\pi e^2}{\hbar} \Theta(t) \sum\limits_{m\neq n} f_{nm} \omega_{nm} \; (g_{\mu\nu}^{nm} + i \Omega^{nm}_{\mu\nu}/2) e^{ i \omega_{mn} t} 
\end{eqnarray}
which we can further simplify as
\begin{equation}
    \sigma_{\mu\nu}(t) = \dfrac{2\pi e^2}{\hbar} \Theta(t)  \sum\limits_{m\neq n} f_{nm} \omega_{nm} \big[ g_{\mu\nu}^{nm}  \cos(\omega_{mn} t) - \Omega^{nm}_{\mu\nu} \sin(\omega_{mn} t)/2  \big] \label{eq:def-sigma-mu-nu-time-domain}
\end{equation}
so that it is explicit that $\sigma_{\mu\nu}(t)$ is a real number. The equation has a consistent structure with symmetric part $g_{mn}$ appearing with symmetric $\cos(\omega_{mn}t)$ and vice versa. 
We can make an attempt to extract the band resolved quantum geometry
\begin{equation}
    \sigma_{\mu\nu}(t) = \dfrac{2\pi e^2}{\hbar} \Theta(t)  \sum\limits_{n} f_{n} \left(  \sum\limits_{m\neq n} \omega_{nm} \big[ g_{\mu\nu}^{nm}  \cos(\omega_{mn} t) - \Omega^{nm}_{\mu\nu} \sin(\omega_{mn} t)/2  \big] \right) \label{eq:def-sigma-mu-nu-time-domain-band-resolved}
\end{equation}
but the term inside the bracket is off by a factor of $\omega_{mn}$, in comparison to the definition of quantum metric and Berry curvature in Eq.~\eqref{eq:g-omega-decomposition-of-xmu-xnu}.
This motivates us to endow time-dependence to quantum geometry that may potentially induce factors of $\omega_{mn}$.
We will explore this in the next section.

\begin{comment}
Before closing this section, we note an important identity
\begin{equation}
    \sum\limits_{n\neq m} f_{nm} \mathcal{T}_{mn} = \sum\limits_{n\neq m} f_n(1-f_m) ( \mathcal{T}_{mn}- \mathcal{T}_{nm}) = \sum\limits_{n} f_n \left( \sum\limits_{m\neq n} (\mathcal{T}_{mn}- \mathcal{T}_{nm})\right)\label{eq:trick:nm-swap}
\end{equation}
that is true for any tensor $\mathcal{T}_{mn}$.
\end{comment}

\section{Time Dependent Quantum Geometric Tensor}\label{app:sec:tQGT}
\rev{Quantum geometry is related to the Riemannian geometry of a projected manifold. It thus requires the notion of a projected manifold, defined by $\hat{P}$, and its complementary manifold, characterized by $\hat{Q} = 1 - \hat{P}$.
In the context of dipole transitions relevant to conductivity and sum rules, it is better to express the dynamics in terms of the off-diagonal dipole operator, $\mathcal{D}_\mu(t) = \hat{Q} \hat{r}_\mu(t)\hat{P}$, which mediates transitions between the occupied and unoccupied states.
The time-dependent Quantum Geometric Tensor (tQGT) can be formulated as the correlation function
\begin{equation}
    \mathcal{Q}_{\mu\nu}(t - t^\prime) = \left\langle  \mathcal{D}^\dag_\mu(t) \mathcal{D}_\nu(t^\prime) \right\rangle = \left\langle  \hat{P} \hat{r}_\mu(t) \hat{Q} \hat{r}_\nu(0) \right\rangle \label{eq:def:introduce-Q-munu-t}
\end{equation}
under the assumption that $\hat{P}$ and $\hat{Q}$ respectively project onto filled and empty states. This assumption is true at zero temperature, where each quantum state is either fully occupied or fully empty, meaning the occupation factor $f_m$ is either $0$ or $1$. 
However, with partial occupations ($0 < f_m < 1$) the distinction between the occupied and empty manifolds becomes less clear, leading to overlap between the two. Such cases require more careful treatment. Therefore, unless otherwise specified, we will take the zero-temperature value for all occupation factors in our discussion. A detailed discussion of the finite temperature framework is provided in Sec.~\ref{app:finite:temperature}.
}

As noted in the main text, tQGT describes correlation between dipole moments mediated entirely via virtual states.
We next use the decomposition of position operators from Eq.~\eqref{eq:g-omega-decomposition-of-xmu-xnu} to write it as
\begin{equation}
    \mathcal{Q}_{\mu\nu}(t) =  \sum\limits_{m,n} f_n(1-f_m) e^{i \omega_{mn} t} \; (g_{\mu\nu}^{mn} + i \Omega_{\mu\nu}^{mn}/2) \label{eq:def:Qmunu-g-omega}.
\end{equation}

The tensor has several interesting properties. At $t=0$, it is equal to the quantum geometric tensor whose real and imaginary parts are the quantum metric and Berry curvature
\begin{equation}
    \mathcal{Q}_{\mu\nu}(t=0) = \left(  \sum\limits_{m, n} f_n(1-f_m) \;g_{\mu\nu}^{mn} \right) + \dfrac{i}{2} \left(  \sum\limits_{m,n} f_n(1-f_m)\Omega_{\mu\nu}^{mn} \right).
\end{equation}
At finite $t$, the presence of $\hat{Q}$ destroys the Hermiticity with $\mathcal{Q}_{\mu\nu}(t)^\dag = \mathcal{Q}_{\nu\mu}(-t) \neq \mathcal{Q}_{\mu\nu}(t)$. 
However, we can always split it into Hermitian and anti-Hermitian components
\begin{equation}
    \mathcal{Q}_{\mu\nu}(t) = \dfrac{1}{2} (\mathcal{Q}_{\mu\nu}(t) + \mathcal{Q}_{\nu\mu}(-t) ) + \dfrac{1}{2} (\mathcal{Q}_{\mu\nu}(t) - \mathcal{Q}_{\nu\mu}(-t) ) \equiv \mathcal{Q}^{\rm s}_{\mu\nu}(t) - \dfrac{i}{2} \mathcal{Q}^{\rm as}_{\mu\nu}(t)
\end{equation}
which introduces Hermitian tensors 
\begin{eqnarray}
    \mathcal{Q}^{\rm s}_{\mu\nu}(t) &=&  \sum\limits_{m, n} f_n(1-f_m) \big[ \cos(\omega_{mn}t ) g_{\mu\nu}^{mn} -  \sin(\omega_{mn}t ) \Omega_{\mu\nu}^{mn}/2 \big] \\
    \mathcal{Q}^{\rm as}_{\mu\nu}(t) &=&  \sum\limits_{m, n} f_n(1-f_m) \big[ 2\sin(\omega_{mn}t ) g_{\mu\nu}^{mn} +  \cos(\omega_{mn}t ) \Omega_{\mu\nu}^{mn} \big].
\end{eqnarray}
These quantities oscillate between quantum geometry and Berry curvature. More importantly, they reformulate the zero point dynamics of charge to the presence of quantum metric and Berry curvature.

Returning to our goal of making connections with conductivity, we note that $\mathcal{Q}^{\rm as}_{\mu\nu}$ can be rewritten as
\begin{equation}
    \mathcal{Q}^{\rm as}_{\mu\nu}(t) =  2\sum\limits_{m\neq n} f_{nm} \big[ \sin(\omega_{mn}t ) g_{\mu\nu}^{mn} +  \cos(\omega_{mn}t ) \Omega_{\mu\nu}^{mn}/2 \big] \label{eq:def-F-mu-nu}
\end{equation}
so that upon taking a time derivative
\begin{equation}
    \partial_t\mathcal{Q}^{\rm as}_{\mu\nu}(t) =  2\sum\limits_{m\neq n} f_{nm} \omega_{mn} \big[ \cos(\omega_{mn}t ) g_{\mu\nu}^{mn} -  \sin(\omega_{mn}t ) \Omega_{\mu\nu}^{mn}/2 \big]
\end{equation}
we get an expression that is identical to conductivity in the time domain in Eq.~\eqref{eq:def-sigma-mu-nu-time-domain}. This leads us to the exact relation
\begin{equation}
    \sigma_{\mu\nu}(t) = \dfrac{\pi e^2}{\hbar} \;\Theta(t)\; \partial_t \mathcal{Q}^{\rm as}_{\mu\nu}(t) .\label{eq:sigma-Q-relation}
\end{equation}

\section{Time Dependent Quantum Geometry in Crystalline solids}\label{app:sec:tQGT-in-solids}
\rev{

The many-body formula for conductivity in Eq.~\eqref{app:eq:def-sigma} is written in terms of the exact many-body energies $\{ E_m \}$ and states $|u_m\rangle$ in the micro-canonical ensemble with fixed number of particles.
In this section, we focus on non-interacting electrons in crystalline solids, which simplifies things considerably.
First, the many-body state can be described using Slater determinants of single-particle states, which are governed by Bloch's theorem. These single-particle states yield energies $\epsilon_{n, {\bf k}}$ labelled by band $n$ and crystal momentum ${\bf k}\; \in \; {\rm BZ}$. Second, we extend the ensemble from micro-canonical to grand-canonical by introducing chemical potential $\mu$. 
Finally, since both position and current operators conserve particle number, their matrix elements are zero between states with different particle numbers. These simplifications reduce the many-body matrix elements in Eq.~\eqref{app:eq:def-sigma} to those between single-particle states, involving a sum over internal labels like band and crystal momenta. As a result,}
the quantum geometric tensor in Eq.~\eqref{eq:def:introduce-Q-munu-t} can then be expressed in terms of cell-periodic parts of the Bloch wavefunctions $\{| u_{m, {\bf k}^\prime}\rangle\}$ and is given by
\begin{equation}
    \mathcal{Q}_{\mu\nu}(t) = \intBZ d{\bf k} d{\bf k}^\prime \sum\limits_{m,n} f^0_{n, {\bf k}} (1- f^0_{m, {\bf k}^\prime})\; \langle u_{ n, {\bf k} } | \hat{r}_\mu | u_{ m, {\bf k}^\prime } \rangle \langle u_{ m, {\bf k}^\prime } | \hat{r}_\nu | u_{ n, {\bf k} } \rangle\; e^{ i (\omega_{m, {\bf k}^\prime} - \omega_{n, {\bf k}})t } \label{eq:QGT-band-space}
\end{equation}
where the matrix elements of the position operator have a definite structure \cite{Blount1962, Sipe2000}
\begin{equation}
    \langle u_{ n, {\bf k} } | \hat{r}_\mu | u_{ m, {\bf k}^\prime } \rangle = \delta_{mn} \big( -\hbar \delta({\bf k}- {\bf k}^\prime) \langle u_{ n, {\bf k} } | i \partial_\mu u_{ m, {\bf k}^\prime } \rangle + i\hbar \partial_\mu \delta({\bf k}- {\bf k}^\prime) \big) + (1-\delta_{mn}) \delta({\bf k}- {\bf k}^\prime) \langle u_{ n, {\bf k} } | i\hbar \partial_\mu u_{ m, {\bf k}^\prime } \rangle \label{eq:blount-matrix-element}
\end{equation}
\rev{and the Fermi occupation is
\begin{equation}
    f^0_{n, {\bf k}} = \Theta(\mu - \epsilon_{n, {\bf k}})
\end{equation}
to ensure that partial occupations are not allowed for a single state with labels $n, {\bf k}$.} 
We find that the resulting tQGT has two contributions
\begin{equation}
    \mathcal{Q}_{\mu\nu}(t) = \mathcal{Q}^{\rm inter}_{\mu\nu}(t) + \mathcal{Q}^{\rm intra}_{\mu\nu}(t)  
\end{equation}
with the superscripts \rev{``inter'' and ``intra'' refer to inter-band and intra-band terms. As we outline below, both these terms are gauge invariant.}

\subsection{Inter-Band Contribution}\label{app:subsec:inter-band-tQGT}
Inter-band position matrix elements have a delta function in momentum that makes $\langle u_{ n, {\bf k} } | i\partial_\mu u_{ m, {\bf k} } \rangle$ momentum diagonal and we get
\begin{equation}
    \mathcal{Q}^{\rm inter}_{\mu\nu} (t) = \intBZ \sum\limits_{m\neq n}  d{\bf k} f^0_{n, {\bf k}} (1- f^0_{m, {\bf k}})\;  \langle u_{ n, {\bf k} } | i\partial_\mu u_{ m, {\bf k} } \rangle \langle u_{ m, {\bf k} } | i\partial_\nu u_{ n, {\bf k} } \rangle\; e^{ i \omega_{mn, {\bf k}} t } 
\end{equation}
which is identical to Eq.~\eqref{eq:def:Qmunu-g-omega}, with the additional band labels and an integral over the BZ.

\subsection{Intra-Band Contribution}\label{app:subsec:intra-band-tQGT}
\rev{
We will now demonstrate that the intra-band component of the time-dependent Quantum Geometric Tensor (tQGT) corresponds to a contribution from the Fermi surface and leads to the Drude weight.
To begin, observe that the first term in the position matrix element in Eq.~\eqref{eq:blount-matrix-element} is diagonal in both the band indices $(\delta_{m,n})$ and the momenta $(\delta({\bf k} - {\bf k}^\prime))$. As a result, it is Pauli blocked, since the factor $f^0_{m,{\bf k}}(1 - f^0_{m,{\bf k}})$ vanishes identically at zero temperature.
Furthermore, as we will see in Sec.~\ref{app:finite:temperature}, this factor does not contribute to any observable even at finite temperatures.

We substitute one of the position operators in the time-dependent Quantum Geometric Tensor (tQGT) with the matrix element from Eq.~\eqref{eq:blount-matrix-element}, leading to
\begin{equation}
    \mathcal{Q}^{\rm intra}_{\mu\nu}(t) = i \hbar \int_{\text{BZ}} d{\bf k} d{\bf k}^\prime \sum_{m} f^0_{m, {\bf k}} (1 - f^0_{m, {\bf k}^\prime})\; \big( \partial_\mu \delta({\bf k} - {\bf k}^\prime) \big) \; \langle u_{m, {\bf k}^\prime} | \hat{r}_\nu | u_{m, {\bf k}} \rangle\; e^{ i (\omega_{m, {\bf k}^\prime} - \omega_{m, {\bf k}})t } \label{eq:QGT-band-space}.
\end{equation}
It is important to note that the time dependence of this term originates from the exponential factor. In the case of a flat band, where $\omega_{m, {\bf k}} = \omega_{m, {\bf k}^\prime}$, the time dependence vanishes, rendering this term irrelevant for the conductivity, which depends on the time derivative.
Since our primary goal is to compute the Drude weight, we must instead focus on dispersive bands, where $\omega_{m, {\bf k}} \neq \omega_{m, {\bf k}^\prime}$.
This paves the way to use an alternate expression of the position operator in terms of the current operator
\begin{equation}
    \langle u_{ m, {\bf k}^\prime } | \hat{r}_\nu | u_{ m, {\bf k} } \rangle = \dfrac{\langle u_{ m, {\bf k}^\prime } | \hat{J}_\nu | u_{ m, {\bf k} } \rangle}{\omega_{m, {\bf k}} - \omega_{m, {\bf k}^\prime}}, \quad {\bf k}\neq {\bf k}^\prime.
\end{equation}
which would otherwise be highly singular for a flat band.
We now have
\begin{equation}
    \mathcal{Q}^{\rm intra}_{\mu\nu}(t) = i \hbar \int_{\text{BZ}} d{\bf k} d{\bf k}^\prime \sum_{m} f^0_{m, {\bf k}} (1 - f^0_{m, {\bf k}^\prime}) \big( \partial_\mu \delta({\bf k} - {\bf k}^\prime) \big) \dfrac{\langle u_{m, {\bf k}^\prime} | \hat{J}_\nu | u_{m, {\bf k}} \rangle}{\omega_{m, {\bf k}} - \omega_{m, {\bf k}^\prime}}\; e^{i \omega_{m, {\bf k}^\prime, {\bf k}} t},
\end{equation}
where $\omega_{m, {\bf k}^\prime, {\bf k}} \equiv \omega_{m, {\bf k}^\prime} - \omega_{m, {\bf k}}$.
Recognizing that both $\partial_\mu \delta({\bf k} - {\bf k}^\prime)$ and $\omega_{m, {\bf k}^\prime, {\bf k}}$ are odd under the exchange of ${\bf k}$ and ${\bf k}^\prime$, we can decompose the remaining terms into their even and odd components:
\begin{eqnarray}
    f^0_{m, {\bf k}} (1 - f^0_{m, {\bf k}^\prime}) &=& \frac{1}{2} \left( f^0_{m, {\bf k}} + f^0_{m, {\bf k}^\prime} - 2 f^0_{m, {\bf k}} f^0_{m, {\bf k}^\prime} \right) + \frac{1}{2} (f^0_{m, {\bf k}} - f^0_{m, {\bf k}^\prime}), \\
    \langle u_{m, {\bf k}^\prime} | \hat{J}_\nu | u_{m, {\bf k}} \rangle e^{i \omega_{m, {\bf k}^\prime, {\bf k}} t} &=& \frac{1}{2} \left( \langle u_{m, {\bf k}^\prime} | \hat{J}_\nu | u_{m, {\bf k}} \rangle e^{i \omega_{m, {\bf k}^\prime, {\bf k}} t} + \langle u_{m, {\bf k}} | \hat{J}_\nu | u_{m, {\bf k}^\prime} \rangle e^{-i \omega_{m, {\bf k}^\prime, {\bf k}} t} \right) \nonumber \\
    &+& \frac{1}{2} \left( \langle u_{m, {\bf k}^\prime} | \hat{J}_\nu | u_{m, {\bf k}} \rangle e^{i \omega_{m, {\bf k}^\prime, {\bf k}} t} - \langle u_{m, {\bf k}} | \hat{J}_\nu | u_{m, {\bf k}^\prime} \rangle e^{-i \omega_{m, {\bf k}^\prime, {\bf k}} t} \right).
\end{eqnarray}
For the integral to yield a finite result, it must have even parity. This leaves two options: combining the odd parts of the Fermi factors and the current operator, or the even parts of both. We find that only the odd combination gives a time dependent piece. We are thus left with}
\begin{equation}
       \mathcal{Q}^{\rm intra}_{\mu\nu}(t)   =  i\intBZ  d{\bf k} d{\bf k}^\prime \sum\limits_{m} \dfrac{f^0_{m, {\bf k}} - f^0_{m, {\bf k}^\prime}}{\omega_{m, {\bf k}} - \omega_{m, {\bf k}^\prime}} \partial_\mu \delta({\bf k}- {\bf k}^\prime) \Big( \langle u_{ m, {\bf k}^\prime } | \hat{J}_\nu | u_{ m, {\bf k} } \rangle\; \; e^{ i \omega_{m, {\bf k}^\prime, {\bf k}}t } - \langle u_{ m, {\bf k} } | \hat{J}_\nu | u_{ m, {\bf k}^\prime } \rangle \; e^{ -i \omega_{m, {\bf k}^\prime, {\bf k}}t } \Big) \label{eq:Q-intra-exp1}.
\end{equation}
The next step is to use integration by parts in the delta function
\begin{equation}
    \int d{\bf x} \;\partial_\mu \delta({\bf x} - {\bf x}_0) \; f({\bf x}) = -\int d{\bf x} \;\delta({\bf x} - {\bf x}_0) \; \partial_\mu f({\bf x}) = - \partial_\mu f({\bf x}_0).
\end{equation}
The trick here is to note that the term inside the big bracket in Eq.~\eqref{eq:Q-intra-exp1} vanishes when ${\bf k}={\bf k}^\prime$.
The only term that survives after the integration by parts is one that has derivatives of the phases $e^{ -i \omega_{m, {\bf k}^\prime, {\bf k}} t }$.
With this insight, we get
\begin{subequations}
    \begin{align}
       \mathcal{Q}^{\rm intra}_{\mu\nu}(t)  & = i 
       \sum\limits_{m} \left(\lim\limits_{ {\bf k}\rightarrow {\bf k}^\prime } \dfrac{f_{m, {\bf k}} - f_{m, {\bf k}^\prime}}{\omega_{m, {\bf k}} - \omega_{m, {\bf k}^\prime}} \right) \langle u_{ m, {\bf k} } | \hat{J}_\nu | u_{ m, {\bf k} } \rangle \Big( \partial_\mu e^{ i (\omega_{m, {\bf k}^\prime} - \omega_{m, {\bf k}} )t } - \partial_\mu e^{ -i (\omega_{m, {\bf k}^\prime} - \omega_{m, {\bf k}} )t } \Big)_{ {\bf k} = {\bf k}^\prime } \\
       &= t   \sum\limits_{m} \left(-\dfrac{\partial f}{\partial \omega} \right)_{\omega = \omega_{m, {\bf k}}} \partial_\nu \omega_{m,{\bf k}} \; \partial_\mu \omega_{m,{\bf k}} \equiv D_{\mu\nu} t
    \end{align}
\end{subequations}
where we have defined Drude weight $D_{\mu\nu}$ as
\begin{equation}
    D_{\mu\nu} = \sum\limits_{m} \left(-\dfrac{\partial f}{\partial \omega} \right)_{\omega = \omega_{m, {\bf k}}} \partial_\nu \omega_{m,{\bf k}} \; \partial_\mu \omega_{m,{\bf k}} \propto N(0) \langle v_F^2 \rangle_{\rm FS}
\end{equation}
where $N(0)$ is the density of states at the Fermi level and $v_F$ is the Fermi velocity averaged $\langle \cdot \rangle$ over the Fermi surface.

\subsection{Relation to Conductivity and Drude peak}\label{app:subsec:sigma-t-tQGT-drude}
With all terms in place, the tQGT is
\begin{equation}
    \mathcal{Q}_{\mu\nu}(t) = \intBZ \sum\limits_{m\neq n} f^0_{n, {\bf k}} (1- f^0_{m, {\bf k}})\;  \hat{r}_\mu^{nm} \hat{r}_\nu^{mn} \; e^{ i\omega_{m n, {\bf k}} t }  + D_{\mu\nu} t 
\end{equation}
As we showed earlier, conductivity is related to the anti-symmetric part
\begin{equation}
     \mathcal{Q}^{\rm as}_{\mu\nu}(t) = i(\mathcal{Q}_{\mu\nu}(t) - \mathcal{Q}_{\nu\mu}(-t)) = i D_{\mu\nu} t + \intBZ \sum\limits_{m\neq n} f^0_{n, {\bf k}}(1-f^0_{m, {\bf k}}) \big[ \sin(\omega_{mn,{\bf k}}t ) g_{\mu\nu}^{mn}({\bf k}) +  \cos(\omega_{mn,{\bf k}}t ) \Omega_{\mu\nu}^{mn}({\bf k})/2 \big]
\end{equation}
and in particular, conductivity is given by
\begin{eqnarray}
    \sigma_{\mu\nu}(t) &=& \dfrac{2\pi e^2}{\hbar} \Theta(t)  \partial_t \mathcal{Q}^{\rm as}_{\mu\nu} \\
    &=& \dfrac{2\pi e^2}{\hbar} \Theta(t) \left(  2 i  D_{\mu\nu} + \intBZ \sum\limits_{m\neq n} f^0_{n, {\bf k}} (1-f^0_{m, {\bf k}}) \omega_{mn, {\bf k}} \big[ \cos(\omega_{mn, {\bf k}}t ) g_{\mu\nu}^{mn}({\bf k}) - \sin(\omega_{mn, {\bf k}}t ) \Omega_{\mu\nu}^{mn}({\bf k})/2 \big] \right)
\end{eqnarray}
which is identical to the expression for gapped systems, except for the Fermi-surface term, $D_{\mu\nu}$, and the inclusion of the momentum label ${\bf k}$.

\rev{
\subsection{Finite Temperatures}\label{app:finite:temperature}
At finite temperatures, the projector must be replaced by the density matrix $\hat{\rho}$, and the complementary projector by $1 - \hat{\rho}$, which are the correct descriptors for occupation of states. Two key differences arise from this modification. First, the Quantum Geometric Tensor (QGT) becomes
\begin{equation}
    \mathcal{Q}_{\mu\nu}(t - t^\prime) = \tr[\hat{\rho} \;\hat{r}_\mu(t) \;(1 - \hat{\rho}) \;\hat{r}_\nu(0)] 
\end{equation}
where the occupation factors are now allowed to be fractional. Second, the tQGT for non-interacting electrons derived earlier acquires an additional term
\begin{equation}
    \widetilde{\mathcal{Q}}^{\rm intra}_{\mu\nu}(t) = - \hbar \int_{\text{BZ}} d{\bf k} \sum_m f^0_{m, {\bf k}} (1 - f^0_{m, {\bf k}}) \mathcal{A}^\mu_m({\bf k}) \mathcal{A}^\nu_m({\bf k}),
\end{equation}
where $\mathcal{A}^\nu_m({\bf k}) = \langle u_{m, {\bf k}} | i \partial_\mu u_{m, {\bf k}} \rangle$ is the Abelian Berry connection.
Physically, this term arises due to partial occupations, where the overlap between occupied and unoccupied states, $\hat{\rho}(1 - \hat{\rho}) \neq 0$, prevents the projected manifold from coinciding with the manifold of filled states.
Lastly, we note that although this term is explicitly gauge-dependent, it is time-independent and symmetric in $\mu$ and $\nu$, and therefore cancels out in the anti-symmetric part of the tQGT. As a result, it does not contribute to any observable, leaving relations such as Eq.~\eqref{eq:sigma-Q-relation} and all the sum rules unaffected.

We note that notions of finite temperature quantum geometry need symmetric log derivatives of the density matrices that manifestly remove all gauge dependent quantities \cite{ji2025density}.
}

\section{Conductivity in Axial Vector Notation }\label{app:sec:axial}
The discussion so far has been very general and most equations have been written down for arbitrary spatial directions $\mu, \nu$.
In the upocoming sections, we will use dissipative parts of conductivity which require specific spatial directions.
We begin by splitting conductivity into symmetric and anti-symmetric parts
\begin{equation}
    \sigma_{\mu\nu}(\omega) = \dfrac{1}{2}( \sigma_{\mu\nu}(\omega) + \sigma_{\nu\mu}(\omega) ) + \left(\dfrac{1}{2}( \sigma_{\mu\nu}(\omega) - \sigma_{\nu\mu}(\omega) ) \right)
\end{equation}
and then define the axial vectors $\sigma_L$ and $\sigma_H$ to absorb the two
\begin{equation}
    \sigma_{\mu\nu}(\omega) = \delta_{\mu\nu} \sigma_{L,\mu} (\omega) + \epsilon_{\mu\nu\lambda } \sigma_{H,\lambda}(\omega) \label{eq:sigma-e-g-def}.
\end{equation}
In 3D, there are three components to $\sigma_L$ (for $x,y,z$) and three $\sigma_H$ (for $xy,yz,xz$), whereas in 2D, there are two components in $\sigma_L$ but only one in $\sigma_H$ ($xy$).
In the case of an applied magnetic field, $\hat{\lambda}$ is along the direction of the magnetic field.

We now introduce the absorptive part of conductivity
\begin{subequations}
\begin{align}
    \sigma^{\rm abs} &= \dfrac{1}{2} (\sigma_{\mu\nu} + \sigma_{\nu\mu}^*) = \dfrac{1}{2} (\delta_{\mu\nu} \sigma_{L,\mu} (\omega) + \epsilon_{\mu\nu\lambda } \sigma_{H,\lambda}(\omega) + \delta_{\mu\nu} \sigma_{L,\mu}^* (\omega) - \epsilon_{\mu\nu\lambda } \sigma_{H,\lambda}^*(\omega)) \\
    &= \delta_{\mu\nu} \dfrac{1}{2} (\sigma_{L,\mu} (\omega) + \sigma_{L,\mu}^* (\omega) + \epsilon_{\mu\nu\lambda} \dfrac{1}{2} (\sigma_{H,\lambda} (\omega) - \sigma_{L,\lambda}^*  (\omega) ) \\
    &= \delta_{\mu\nu}  {\rm Re}[\sigma_{L,\mu}(\omega)] + i \epsilon_{\mu\nu\lambda} {\rm Im}[\sigma_{H,\lambda}(\omega)]
\end{align}
\end{subequations}
which will eventually give rise to sum rules.

Lastly, we introduce axial vectors for position matrix elements
\begin{equation}
    g^{nm}_{\mu\nu} = g^{nm}_{\mu}  \delta_{\mu\nu}, \quad \Omega^{nm}_{\mu\nu} = 2\epsilon_{\mu\nu\lambda} \Omega^{nm}_{\lambda}  
\end{equation}
to facilitate the discussion on sum rules of dissipative response functions in the upcoming sections.

\section{tQGT and Sum Rules of Dissipative Response}\label{app:sec:sum-rules}
We begin by considering the absorptive part of conductivity
\begin{equation}
\sigma^{\rm abs} = \delta_{\mu\nu}{\rm Re}[ \sigma_{L,\mu}(\omega) ] + i\epsilon_{\mu\nu\lambda} {\rm Im}[ \sigma_{H,\lambda}(\omega) ] = \dfrac{\pi e^2}{\hbar}  \sum\limits_{m\neq n} f_{nm} \omega_{mn}  \; ( g_\mu^{mn} \delta_{\mu\nu} + i\epsilon_{\mu\nu\lambda} \Omega_\lambda^{mn}) \; \delta( \omega - \omega_{mn} ) \label{eq:sigma-abs-def}.
\end{equation}
and the generalized sum rules
\begin{equation}
 \mathcal{S}^\eta_{\mu\nu} = \int\limits_0^\infty d\omega \; \dfrac{ \sigma^{\rm abs}_{\mu\nu}(\omega) ]}{ \omega^{1-\eta}} = \dfrac{\pi e^2}{\hbar}  \sum\limits_{m\neq n} f_{nm} \omega_{mn}  \; ( g_\mu^{mn} \delta_{\mu\nu} + i\epsilon_{\mu\nu\lambda} \Omega_\lambda^{mn}) \; \left[ \int\limits_0^\infty d\omega \; \dfrac{ \delta( \omega - \omega_{mn} )}{ \omega^{1-\eta}} \right]
\end{equation}
where the frequency integral is given by $\Theta(\omega_{mn})/ 2\omega_{mn}^{1-\eta}$. The $\Theta$ function enters because the limits of the integral only access the positive peaks. The resulting sum rule is
\begin{equation}
    \mathcal{S}^\eta_{\mu\nu}  = \dfrac{\pi e^2}{\hbar}  \sum\limits_{m\neq n} f_{nm} \omega_{mn}^\eta  \; ( g_\mu^{mn} \delta_{\mu\nu} + i \epsilon_{\mu\nu\lambda} \Omega_\lambda^{mn}) \; \Theta(\omega_{mn}).
\end{equation}
To link this sum rule with the quantum geometric tensor, we refine the expression by writing the occupation factor $f_{nm}$ as $f_{nm} = f_{n}(1 - f_{m}) - f_{m}(1 - f_{n})$.
We then split the sum into two parts and swap $m\leftrightarrow n$ in the second sum. 
After combining the two we get
\begin{equation}
 \dfrac{\pi e^2}{\hbar}  \sum\limits_{m\neq n} f_{n}(1-f_m)  \omega_{mn}^\eta \big( ( g_\mu^{mn} \delta_{\mu\nu} + i \epsilon_{\mu\nu\lambda} \Omega_\lambda^{mn}) \Theta(\omega_{mn}) - (-1)^\eta  ( g_\mu^{mn} \delta_{\mu\nu} + i \epsilon_{\mu\nu\lambda} \Omega_\lambda^{mn}) \; \Theta(-\omega_{mn}) \big)
\end{equation}
which is best expressed in longitudinal and Hall parts separately
\begin{subequations}
\begin{align}
    \mathcal{S}^\eta_{L,\mu} &= \dfrac{\pi e^2}{\hbar}  \sum\limits_{m\neq n} f_{n}(1-f_m) \omega_{mn}^\eta g^{mn}_\mu \big(  \Theta(\omega_{mn}) - (-1)^\eta \Theta(-\omega_{mn}) \big) \\
    \mathcal{S}^\eta_{H,\lambda} &= \dfrac{\pi e^2}{\hbar}  \sum\limits_{m\neq n} f_{n}(1-f_m) \omega_{mn}^\eta \Omega_\lambda^{mn} \big(  \Theta(\omega_{mn}) + (-1)^\eta \Theta(-\omega_{mn}) \big).
\end{align}
\end{subequations}

The $\Theta(\omega_{mn})$ factors select only half of the resonances. This is crucial for sum rules whose integrand is odd in frequency, which would otherwise cancel if integrated over the full $-\infty$ to $\infty$ range. 
Taking advantage of the gapped spectrum, at $T=0$, with $f_n = 1$ for filled states and $0$ otherwise, we naturally obtain positive $\omega_{mn}$. This action forbids us from using $m\leftrightarrow n$ tricks in the future but simplifies the sum rule to the concise expression
\begin{equation}
    \mathcal{S}^\eta_{L,\mu} = \dfrac{\pi e^2}{\hbar} \sum\limits_{m\neq n} f_{n}(1-f_m)  \omega_{mn}^\eta g_\mu^{mn} ,\quad  \mathcal{S}^\eta_{H,\lambda} = \dfrac{\pi e^2}{\hbar} \sum\limits_{m\neq n} f_{n}(1-f_m) \omega_{mn}^\eta \Omega_\lambda^{mn},
\end{equation}
which we will now relate to the geometric tensor. We first need to extend the $L$ and $H$ terminology to the geometric tensor. We define
\begin{equation}
    \mathcal{Q}_{\mu\nu}(t) = \sum\limits_{m\neq n} f_n(1-f_m) e^{i \omega_{mn} t} \; ( g_\mu^{mn} \delta_{\mu\nu} + \epsilon_{\mu\nu\lambda} \Omega_\lambda^{mn})
\end{equation}
to finally arrive at 
\begin{equation}
    \mathcal{S}_{L/H,\mu}^\eta = \dfrac{\pi e^2}{\hbar} \left[ (-i \hat{\partial}_t)^\eta \mathcal{Q}_{L/H,\mu}(t) \right]_{t=0}
\end{equation}
which explicitly shows that the time-dependent quantum geometric tensor is a generating function for all sum rules.

\subsection{$\eta = 0$ SWM Sum Rule}
We begin with the simplest observation that $\eta=0$ recovers the SWM sum rule
\begin{equation}
    \mathcal{S}^0_{L,\mu} = \int\limits_0^\infty d\omega \; \dfrac{ {\rm Re}[\sigma_{L,\mu}(\omega)]}{ \omega } = \dfrac{\pi e^2}{\hbar} g_{\mu\mu}, \quad \mathcal{S}^0_{H,\lambda} = \int\limits_0^\infty d\omega \; \dfrac{ {\rm Re}[\sigma_{H,\lambda}(\omega)]}{ \omega } = \dfrac{\pi e^2}{\hbar} \mathcal{C}_\lambda  ,
\end{equation}
where $g_{\mu\mu}$ is the quantum metric and $\mathcal{C}_\lambda$ is the Chern number of the occupied bands. Remarkably, the negative moment can extract the quantum geometry of the system.

\subsection{$\eta = 1$ $f$-Sum Rule}
The $\eta = 1$ sum rule defines two distinct physical quantities: optical mass from longitudinal and magnetic moment from Hall.

\begin{itemize}

\item Longitudinal sum rule defines optical mass
\begin{equation}
    \int\limits_0^\infty d\omega \; {\rm Re}[\sigma_{L, \mu}(\omega)] = \dfrac{\pi e^2}{\hbar}  \sum\limits_{m\neq n} f_{n}(1-f_m)  \omega_{mn} g_\mu^{mn}
\end{equation}
where we can further use the fact that $\omega_{mn} g_{mn}$ is anti-symmetric in $m \leftrightarrow n$ to get
\begin{equation}
    \int\limits_0^\infty d\omega \; {\rm Re}[\sigma_{L, \mu}(\omega)] = \dfrac{\pi e^2}{\hbar}  \sum\limits_{n} f_{n} \;\left(\sum\limits_{m\neq n} \omega_{mn} g_\mu^{mn} \right)
\end{equation}
where the quantity in brackets defines an effective mass for the a band
\begin{equation}
    [\mathcal{M}^{-1}_n]_\mu = \sum\limits_{m\neq n} \omega_{mn} g_\mu^{mn}
\end{equation}
going with analogy that the $f$-sum rule gives the plasma frequency that takes the form $n/m^*$ with some effective mass $m^*$. Here we are finding that $m^*$ is indeed related to the quantum geometry.
Moreover, the mass can be resolved for each state $n$.

Lastly, we note that in non-interacting metals, the Fermi-surface modifies the effective mass with the band curvature as we saw already in sec.~\ref{app:subsec:intra-band-tQGT}.
The additional contribution is the Drude piece
\begin{equation}
    D_{\mu\nu} = \intBZ d{\bf k} \sum\limits_{m} \left(-\dfrac{\partial f^0}{\partial \omega} \right)_{\omega = \omega_{m, {\bf k}}} \partial_\nu \omega_{m,{\bf k}} \; \partial_\mu \omega_{m,{\bf k}} = \intBZ d{\bf k} \sum\limits_{m} f^0_{m, {\bf k}} \partial^2_{\mu\nu} \epsilon_{m,{\bf k}}
\end{equation}
which gives the full sum rule
\begin{equation}
    \int\limits_0^\infty d\omega \; {\rm Re}[\sigma_{L, \mu}(\omega)] = \dfrac{\pi e^2}{\hbar} \intBZ d{\bf k} \sum\limits_{m} f^0_{m, {\bf k}} \big( \partial^2_{\mu} \epsilon_{m,{\bf k}} + [\mathcal{M}^{-1}_m({\bf k})]_\mu \big)
\end{equation}
which was used in ref.~\cite{Hazra2019} to get upper bounds on superfluid stiffness in multi-band systems.

\item Hall sum rule defines magnetic moment
\begin{equation}
    \int\limits_0^\infty d\omega \; {\rm Im}[\sigma_{H,\lambda}(\omega)] = \dfrac{\pi e^2}{\hbar}  \sum\limits_{m\neq n} f_{n}(1-f_m)  \omega_{mn} \Omega_\lambda^{mn}
\end{equation}
where 
\begin{equation}
    \mu^\lambda =  \sum\limits_{ \substack{n\in {\rm filled}\\ m \in {\rm empty} } }\omega_{mn} \Omega_\lambda^{mn}
\end{equation}
is defined to be the net orbital magnetic moment of the filled bands (see eq.~(12) in Ref.~\cite{Souza2008}).
     
\end{itemize}

\subsection{$\eta = -1$ Dielectric Permittivity}
The longitudinal response
\begin{equation}
    \int\limits_0^\infty d\omega \; \dfrac{ {\rm Re}[\sigma_{L,\mu}(\omega)]}{ \omega^2 } = \dfrac{\pi e^2}{\hbar}  \sum\limits_{m\neq n} f_{n}(1-f_m) \dfrac{ g_\mu^{mn} }{ \omega_{mn} } = \dfrac{\pi e^2}{\hbar}  \sum\limits_{m\neq n} f_{n} \left( \sum\limits_{ m\neq n} \dfrac{ g_\mu^{mn} }{ \omega_{mn} } \right)
\end{equation}
defines the electric susceptibility of the $n^{\rm th}$ band as
\begin{equation}
    \chi_\mu^{e,n} = \sum\limits_{ \substack{n\in {\rm filled}\\ m \in {\rm empty} } } \dfrac{ g_\mu^{mn} }{ \omega_{mn} }
\end{equation}
related to the capacitance by the vacuum permittivity $c_\mu^{e,n}=\epsilon_0\chi_\mu^{e,n}$. Similarly, the Hall response is obtained by the nonreciprocal part of the sum rule, only present if time-reversal symmetry is broken
\begin{equation}
    \int\limits_0^\infty d\omega \; \dfrac{ {\rm Im}[\sigma_{H,\lambda}(\omega)]}{ \omega^2 } = \dfrac{\pi e^2}{\hbar}  \sum\limits_{m\neq n} f_{n}(1-f_m) \dfrac{ \Omega_\lambda^{mn} }{ \omega_{mn} }
\end{equation}
defines the quantity
\begin{equation}
    \chi^\lambda_m = \sum\limits_{ \substack{n\in {\rm filled}\\ m \in {\rm empty} } }  \dfrac{ \Omega_\lambda^{mn}}{ \omega_{mn} }.
\end{equation}

\section{Bounds on Sum Rules}\label{app:sec:bounds}
Following ref.~\cite{traini1996electric}, we consider a function of generalized sum rules of longitudinal conductivity (which is positive definite)
\begin{equation}
    h(\beta) = \int\limits_0^\infty \; \dfrac{ {\rm Re}[\sigma_{L,\mu}(\omega)] }{ \omega }\; ( \omega^m + \beta \omega^n )^2
\end{equation}
which can be written as
\begin{equation}
h(\beta) = S_{2m} + 2 \beta S_{m+n} + \beta^2 S_{2n}
\end{equation}
where we have used the shorthand $S_{m} = \mathcal{S}^m_{L,\mu}$ for the longitudinal sum rule.
Since ${\rm Re}[\sigma_{L,\mu}(\omega)] \geq 0$, and the limits of the integral are from $0$ to $\infty$, the function $h$ is positive semi-definite. It implies that the minimum of $h$ is positive semi-definite as well.
Finding the minima gives the condition
\begin{equation}
    h^\prime(\beta^*) = 0, \quad \beta^* = -\dfrac{S_{m+n}}{ S_{2n} }, \quad h(\beta^*) = S_{2m} - \dfrac{S_{m+n}^2 }{S_{2n}} \geq 0
\end{equation}
and hence we get the relation
\begin{equation}
    S_{m+n}^2 \leq S_{2m} S_{2n}
\end{equation}
which is valid for any $m,n \in \mathbb{R}$.

\section{Tight-Binding Models}
For completeness, we enumerate the details of various tight-binding models used in the main text.

\subsection{Honeycomb Lattice with Inversion Breaking Mass Term}\label{app:subsec:TB-honeycomb-plus-mass}
We consider a honeycomb lattice with two inequivalent sites $A$ and $B$. With one orbital at each site, the lattice is described by the lattice vectors and basis
\begin{equation}
 {\bf a}_1 = (\sqrt{3}/2, 1/2), \quad {\bf a}_2 = (\sqrt{3}/2, -1/2) , \quad \boldsymbol\tau = \dfrac{1}{3}( {\bf a}_1 + {\bf a}_2 )
\end{equation}
with $\tau_A = {\bf 0}$ and $\tau_B = \boldsymbol\tau$.
The Bloch Hamiltonian for nearest neighbor hopping and inversion breaking mass is given by
\begin{equation}
    \mathcal{H} = \intBZ d{\bf k} \; \begin{pmatrix} c^\dag_{A, {\bf k}} & c^\dag_{B, {\bf k}} \end{pmatrix} \begin{pmatrix}
        m_z & \gamma({\bf k}) \\ \gamma({\bf k})^* & -m_z
    \end{pmatrix} \begin{pmatrix} c^\phdag_{A, {\bf k}} \\ c^\phdag_{B, {\bf k}} \end{pmatrix}
\end{equation}
where 
\begin{equation}
    \gamma({\bf k}) = \sum\limits_{j=1}^3 e^{ i {\bf k}\cdot \boldsymbol{\delta}_j}, \quad \boldsymbol{\delta}_j = R_z (2\pi j /3 ) \boldsymbol{\tau}.
\end{equation}
$R_z(\theta)$ here denotes a rotation matrix about the $\hat{z}$ axis by angle $\theta$. The resulting band structure has two bands with the minimum gap given by $\hbar \omega_g = 2 m_z$.

\subsection{Haldane Model for Chern Insulator}\label{app:subsec:Haldane}
Haldane model includes both nearest neighbor hopping and inversion breaking mass as described in the previous section, in addition to a next-nearest neighbor hopping with a flux \cite{Haldane1988}.
The Bloch Hamiltonian is given by
\begin{equation}
    \mathcal{H} = \intBZ d{\bf k} \; \begin{pmatrix} c^\dag_{A, {\bf k}} & c^\dag_{B, {\bf k}} \end{pmatrix} \begin{pmatrix}
        m_z + t_2 \Gamma(\phi, {\bf k}) & t\gamma({\bf k}) \\ t\gamma({\bf k})^* & -m_z + t_2 \Gamma(-\phi, {\bf k})
    \end{pmatrix} \begin{pmatrix} c^\phdag_{A, {\bf k}} \\ c^\phdag_{B, {\bf k}} \end{pmatrix}
\end{equation}
where
\begin{equation}
    \Gamma(\phi, {\bf k}) = 2\sum\limits_{j=1}^3 \cos({\bf k}\cdot {\bf A}_j + \phi), \quad {\bf A}_j = R_z (2\pi j /3 ) {\bf a}_1.
\end{equation}
The model has two bands with gap $m_z \pm 3\sqrt{3} t_2 \sin(\phi)$ at the two valleys $\pm {\bf K}$. The system undergoes a topological phase transition at $m_z = 3\sqrt{3}t_2\sin(\phi)$. For $\phi=\pi/2$, this corresponds to $m_z \approx 5.2 t_2$.

\subsection{Lieb Lattice to Square Lattice Model}\label{app:subsec:lieb-square}
We consider a four-orbital tight-binding model with lattice vectors
\begin{equation}
    {\bf a}_1 = (1,0), \quad {\bf a}_2 = (0,1), \quad \boldsymbol{\tau}_A = {\bf 0}, \quad \boldsymbol{\tau}_B = {\bf a}_1/2, \quad \boldsymbol{\tau}_C = {\bf a}_2/2, \quad \quad \boldsymbol{\tau}_D = {\bf a}_1/2 + {\bf a}_2/2
\end{equation}
given by the Bloch Hamiltonian
\begin{equation}
    \mathcal{H} = \intBZ d{\bf k} \; \begin{pmatrix} c^\dag_{A, {\bf k}} & c^\dag_{B, {\bf k}} & c^\dag_{C, {\bf k}} & c^\dag_{D, {\bf k}} \end{pmatrix} \begin{pmatrix}
        0 & t g_x & t g_y & 0 \\
        t g_x & 0 & 0 & t_p g_y \\
        t g_y & 0 & 0 & t_p g_x \\
        0 & t_p g_y & t_p g_x & 0
    \end{pmatrix} \begin{pmatrix} c^\phdag_{A, {\bf k}} \\ c^\phdag_{B, {\bf k}} \\ c^\phdag_{C, {\bf k}} \\ c^\phdag_{D, {\bf k}} \end{pmatrix}
\end{equation}
where the function $g$ is defined as
\begin{equation}
    g_x = -2 \cos( {\bf k}\cdot {\bf a}_1/2), \quad g_y = -2 \cos( {\bf k}\cdot {\bf a}_2/2).
\end{equation}

When $t_p = t$, the model is identical to a square lattice, expressed in a larger unit cell. With $t_p=0$, the $D$ orbital factors out and the resulting lattice is the Lieb lattice with an exact flat band at zero energy.

\end{document}